\documentclass{article}
\pdfoutput=1
\usepackage[utf8]{inputenc}
\usepackage[sorting=none]{biblatex}
\usepackage{multicol}
\usepackage{graphicx}
\usepackage{appendix}
\usepackage{authblk}
\usepackage[total={6.5in,8.75in},
top=1.2in, left=1.4in, right = 1.4in]{geometry}
\title{Binary Star Population with Common Proper Motion in Gaia DR2}
\author{S. A. Sapozhnikov, D. A. Kovaleva, O. Yu. Malkov, A. Yu. Sytov}
\affil{Instutute of Astronomy RAS, Moscow, Russia}
\date{September 2020}

\begin{document}
\def\apj{Astrophys.~J}
\def\aatr{Astron.~Astroph.~Trans}
\def\aaps{Astron.~and Astrophys.~Suppl.~Ser}
\def\pasp{Publ.~Astron.~Soc.~Pac}
\def\gca{Geochim.~Cosmochim.~Acta}
\def\aap{Astron.~Astrophys}
\def\aspcs{ASP~Conf.~Ser}
\def\asrep{Astron.~Rep}
\def\nat{Nature}
\def\apjl{Astrophys.~J.~Lett}
\def\apjs{Astrophys.~J.~Suppl.~Ser}
\def\aj{Astron.~J}
\def\mnras{Mon.~Not.~R.~Astron.~Soc}
\def\araa{Ann.~Rev.~Astron.~Astrophys}
\def\jcp{J.~Chem.~Phys}
\def\apss{Astrophys.~Space.~Sci}
\def\prl{Phys.~Rev.~Lett}
\def\phrva{Phys.~Rev.~A}
\def\phlb{Phys.~Let.~B}
\def\pf{Phys.~Fluids}
\def\azh{áâà®­.~¦ãà­}
\def\pazh{¨áì¬ ~¢~áâà®­.~¦ãà­}
\def\jgr{J.~Geophys.~Res}
\def\cemda{Celest.~Mech.~Dyn.~Astr}
\def\jcoph{J.~Comp.~Phys}
\def\cophc{Comput.~Phys.~Commun}
\def\phpl{Physics~of~Plasmas}
\def\pasj{Publ.~Astron.~Soc.~Jpn}
\def\avest{áâà®­.~¢¥áâ­}
\def\jrasc{J.~R.~Astron.~Soc.~Can}
\def\cemec{Celest.~Mech}
\def\pasau{Proc.~Astron.~Soc.~Aust}
\def\puasau{Publ.~Astron.~Soc.~Aust}
\def\jasa{J.~Acoust.~Soc.~Am}
\def\jfm{J.~Fluid~Mech}
\def\cajph{Can.~J.~Phys}
\def\mitag{Mitt.~Astron.~Ges}
\def\bain{Bull.~Astron.~Inst.~Neth}
\def\epsl{Earth~Planet.~Sci.~Lett}
\def\ibvs{Inf.~Bull.~Variable~Stars}
\def\arep{Astr.~Rep}
\def\phr{Phys.~Rep}
\def\astl{Astron.~Letters}
\def\sci{Science}
\def\jqsrt{J.~Quant.~Spectrosc.~Radiat.~Transfer}
\def\emp{Earth,~Moon~and~Planets}
\def\icar{Icarus}
\def\pss{Planet.~Space~Sci}
\def\qjras{Q.~J.~R.~Astron.~Soc}
\def\nimpa{Nucl.~Instrum.~Methods~Phys.~Res.,~Sect.~A}
\def\soph{Sol.~Phys}
\def\lnm{Lect.~Notes~in~Math}
\def\an{Astron.~Nach}
\def\aph{Astroparticle~Physics}
\def\adspr{Adv.~Space~Res}
\def\geoj{Geophys.~J}
\def\caosp{Contrib.~Astron.~Obs.~Skalnat{\'e}~Pleso}
\def\vestcpbu{¥áâ­.~.-¥â¥à¡.~ã­-â }
\def\izvvrad{§¢.~¢ã§®¢.~ ¤¨®ä¨§¨ª }
\def\izvans{§¢.~AH~CCCP}
\def\vestvgu{¥áâ­.~®«}
\def\vestsibgau{¥áâ­.~¨¡}
\def\bamass{Bull.~Am.~Astron.~Soc}
\def\rmxaa{Rev.~Mex.~Astron.~Astrofis}
\def\aapr{Astron.~Astrophys.~Rev}
\def\acp{Atmosphere~Chem.~Phys}
\def\cosiss{®á¬¨ç.~¨áá«¥¤}
\def\ssrv{Space~Sci.~Rev}
\def\jmph{J.~Math.~Phys}
\def\rvmps{Rev.~Mod.~Phys.~Suppl}
\def\rvmp{Rev.~Mod.~Phys}
\def\prd{Phys.~Rev.~D}
\def\nuphs{Nuc.~Phys.~B~Proc.~Suppl}
\def\nuphb{Nuc.~Phys.~B}
\def\skytel{Sky~Telesc}
\def\thmc{¥§.~¬¥¦¤ã­ à®¤.~ª®­ä}
\def\mmc{ â¥à¨ «ë~¬¥¦¤ã­ à®¤.~ª®­ä}
\def\mvrc{ â¥à¨ «ë~¢á¥à®á.~ª®­ä}
\def\cntc{¡.~­ ãç.~âà.~ª®­ä}
\def\tmnpc{à.~¥¦¤ã­ à®¤.~­ ãç.-¯à ªâ.~ª®­ä}
\def\ctc{¡.~âà.~ª®­ä}
\def\mcnctone{à.~31-© ¥¦¤ã­ à®¤.~áâã¤.~­ ãç.~ª®­ä., ª â¥à¨­¡ãà£, 28 ï­¢.~--- 1~ä¥¢à. 2002~£}
\def\mcnctto{à.~32-© ¥¦¤ã­ à®¤.~áâã¤.~­ ãç.~ª®­ä., ª â¥à¨­¡ãà£, 3---7 ä¥¢à. 2003~£}
\def\mcncttr{à.~33-© ¥¦¤ã­ à®¤.~áâã¤.~­ ãç.~ª®­ä., ª â¥à¨­¡ãà£, 2---6 ä¥¢à. 2004~£}
\def\mcnctfo{à.~34-© ¥¦¤ã­ à®¤.~áâã¤.~­ ãç.~ª®­ä., ª â¥à¨­¡ãà£, 31 ï­¢.~--- 4~ä¥¢à. 2005~£}
\def\mcnctfi{à.~35-© ¥¦¤ã­ à®¤.~áâã¤.~­ ãç.~ª®­ä., ª â¥à¨­¡ãà£, 30 ï­¢.~--- 3~ä¥¢à. 2006~£}
\def\mcnctsi{à.~36-© ¥¦¤ã­ à®¤.~áâã¤.~­ ãç.~ª®­ä., ª â¥à¨­¡ãà£, 29 ï­¢.~--- 2~ä¥¢à. 2007~£}
\def\mcnctse{à.~37-© ¥¦¤ã­ à®¤.~áâã¤.~­ ãç.~ª®­ä., ª â¥à¨­¡ãà£, 28 ï­¢.~--- 1~ä¥¢à. 2008~£}
\def\mcnctei{à.~38-© ¥¦¤ã­ à®¤.~áâã¤.~­ ãç.~ª®­ä., ª â¥à¨­¡ãà£, 2---6 ä¥¢à. 2009~£}
\def\mcnctni{à.~39-© ¥¦¤ã­ à®¤.~áâã¤.~­ ãç.~ª®­ä., ª â¥à¨­¡ãà£, 1---5 ä¥¢à. 2010~£}
\def\mcncforty{à.~40-© ¥¦¤ã­ à®¤.~áâã¤.~­ ãç.~ª®­ä., ª â¥à¨­¡ãà£, 31 ï­¢.~--- 4~ä¥¢à. 2011~£}
\def\mcncfone{à.~41-© ¥¦¤ã­ à®¤.~áâã¤.~­ ãç.~ª®­ä., ª â¥à¨­¡ãà£, 30 ï­¢.~--- 3~ä¥¢à. 2012~£}
\def\tmc{à.~¥¦¤ã­ à®¤.~ª®­ä}
\def\tc{à.~ª®­ä}
\def\thc{¥§.~ª®­ä}

\def\IAUsympc{Proc.~IAU~Symp}
\def\IAUsymp{Proc.~IAU~Symp}
\def\IAUcoll{Proc.~IAU~Colloquia}
\def\prconf{Proc.~conf}
\def\princonf{Proc.~int.~conf}
\def\mvrnc{ â¥à¨ «ë~¢á¥à®á.~­ ãç.~ª®­ä}
\def\na{New Astronomy}

\maketitle
\begin{abstract}
	We describe a homogeneous catalog compilation of common proper motion stars based on Gaia DR2. A preliminary list of all pairs of stars within the radius of 100 pc around the Sun with a separation less than a parsec was compiled. Also, a subset of comoving pairs, wide binary stars, was selected. The clusters and systems with multiplicity larger than 2 were excluded from consideration. The resulting catalog contains 10358 pairs of stars. The catalog selectivity function was estimated by comparison with a set of randomly selected field stars and with a model sample obtained by population synthesis. The estimates of the star masses in the catalogued objects, both components of which belong to the main-sequence, show an excess of “twins”, composed by stars with similar masses. This excess decreases with increasing separation between components. It is shown that such an effect cannot be a consequence of the selectivity function only and does not appear in the model where star formation of similar masses is not artificially preferred. The article is based on the talk presented at the conference “Astrometry yesterday, today, tomorrow” (Sternberg Astronomical Institute of the Moscow State University, October 14–16, 2019).
\end{abstract}

\section{Introduction}
A significant proportion of the stellar population is
concentrated in the binary stars. According to some
estimates \cite{WideBinariesKepler}, a half of all main-sequence stars are
components of binary and multiple systems. Binaries
are an important component of the Galaxy’s stellar
population, which affects the stellar system evolution
as a whole. In addition, the parameters of stellar pairs
can provide us important information about star formation.
It is generally accepted that the stars are born in
large groups, which eventually decay into separate systems.
The vast majority of binary and multiple stars
are the remainders of such groups. Thus, examining a
multiple system, its components likely formed simultaneously,
under the same conditions \cite{2005A&A...439..565G}. The study of
stellar systems with various characteristics allows one
to advance in solving many astrophysical problems
(see, i.e., the discussion in \cite{2013ARA&A..51..269D}).The dynamic connection
between the components of pairs itself makes it
possible to evaluate some of the physical parameters of
them directly. Interaction between components of
close pairs leads to the formation of various astronomical
objects attractive for the study, e.g., Novae, type Ia
Supernovae, pulsars, and symbiotic binary stars.
In the present paper, we consider the population of
wide binary stars. Thus, the pairs with separation of
components are so large, their evolution proceeds in
the same way as in single stars. Among such binaries,
there are both pairs with observed orbital motion and
pairs with a common proper motion (comoving ones)
with orbital periods from several thousand to millions
of years and with the distance between components up
to many thousands of astronomical units \cite{binariesDR2feb}. Wide
binaries are loosely bound and can be easily destroyed
by the heterogeneities of the Galaxy’s potential, for
instance, due to the close passage of giant molecular
clouds. This makes such systems valuable indicators of
the Galaxy’s dynamic environment. The distribution
of wide binary stars over masses of components and
the distance between them makes it possible to elucidate
the features of star formation processes.
The present study is aimed to the population properties
of wide binaries and common proper motion
stars at the distance of up to 100 pc from the Sun,
identified in the study of the catalog of candidates for
pairs with a common proper motion \cite{Catalogue}, created on
the basis of the Gaia DR2 database.
Section 2 briefly outlines the principles of creating
the catalog and refining it. In Section 3, the selected
population parameters of binary stars with common
proper motion are investigated and discussed. Conclusions
are drawn in Section 4.

\section{The catalog of binary and comoving stars}

\begin{figure}[h]
		\includegraphics[width=0.45\linewidth]{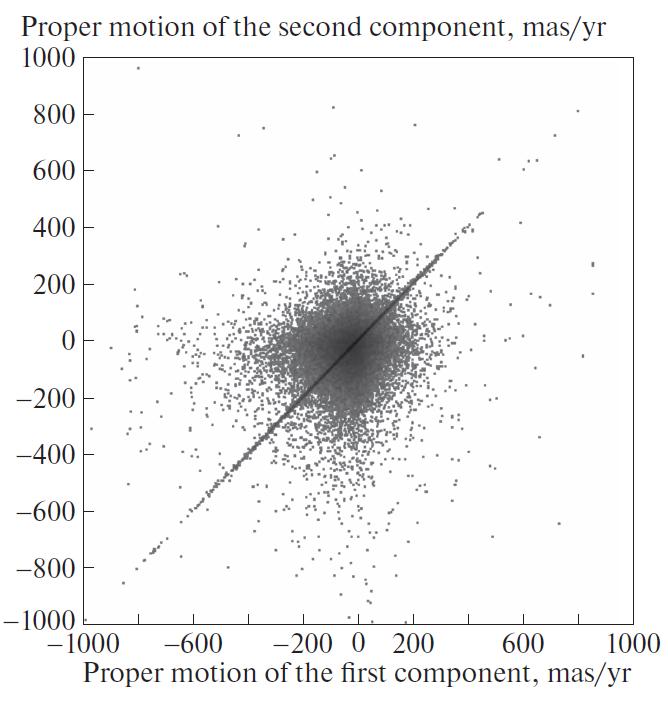}
		\includegraphics[width=0.45\linewidth]{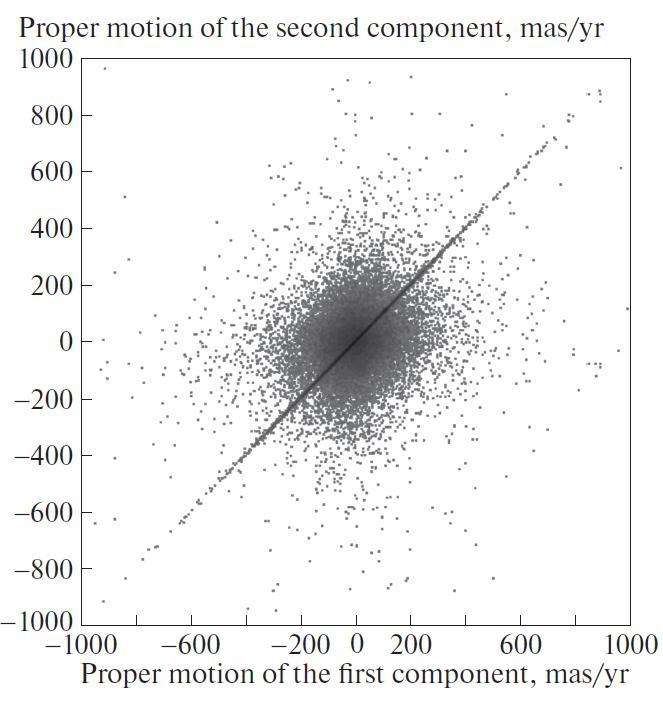}
		\caption{Distribution of the pair components according to their proper motion according to declination (left) and right ascension
(right). Position of the individual points along two axes shows the proper motion of two components. A much densely populated
diagonal is visible, corresponding to the pairs with a common proper motion.}
		\label{fig:pm}
	\end{figure}

Candidate binary stars were selected among the
objects of the Gaia DR2 catalog that have parallaxes $10 \leq \varpi \leq 100$ mas, which correspond, when estimating distance from parallax with $\frac{1000}{\varpi}$ [pc], , to
the stars located inside a spherical layer with an inner
radius of 10 pc and an outer radius of 100 pc around
the Sun. The lower limit of the distance to the Sun is
due to computational constraints related to the procedure
of candidate selection (see Appendix). The upper
limit of distance (100 pc) is selected for a number of
the following reasons. Within this volume, the characteristic
relative parallax error does not exceed $10 \%$, which, allows us to use the parallax value for
estimating distances as described above (see, for
example, \cite{2015PASP..127..994B, 2017AstBu..72..122R}). Besides, Gaia DR2 completeness limit for faint stars is $G \approx 17^m$ \cite{2018A&A...616A...1G}, which means that it
includes all stars of the lower part of the main sequence
within 100 pc, at least up to the spectral subtype
$M4.5$ (such stars, according to Mamajek’s data \footnote{\url{http://www.pas.rochester.edu/~emamajek/EEM_dwarf_UBVIJHK_colors_Teff.txt}}, have an absolute magnitude of $G \approx 12^m$), and within Gaia horizon (up to apparent magnitude of $G \approx 21^m$) all dwarfs up to M10 are located. As potential candidates for binary components, not all
Gaia sources in this field are considered, but only
those satisfying the quality limits of the astrometric
and photometric solutions (see \cite{GDR2_AstrometricSolution}, Gaia DR2 Known Issues\footnote{\url{https://www.cosmos.esa.int/web/gaia/dr2-known-issues},\\ \url{https://www.cosmos.esa.int/documents/29201/1770596/Lindegren_GaiaDR2_Astrometry_extended.pdf/1ebddb25-f010-6437-cb14-0e360e2d9f09}}).
The following restrictions on the astrometric solution
quality were applied: first, the RUWE (Renormalized
Unit Weight Error) parameter should not
exceed 1.4 and, second, the nominal relative parallax
error should not exceed $10\%$. The quality of the photometric
solution was checked by the limit on the Flux
Excess Factor  \cite{GDR2_AstrometricSolution}. The introduction of these filters
results in rejection of many weak sources concentrated
mainly in the direction of the Galactic center with a
density that is much higher than expected from the
notion on the star distribution in the Galaxy. We
assume that these sources may be more distant stars.
Due to the sky background and the high density of
sources in the Galaxy plane, the determination reliability
of parallaxes is low.
	
	40\% of Gaia sources within 100 pc from the Sun
satisfy the above-mentioned restrictions. These
242122 sources are treated as the stars among which
we search for binary and common proper motion
stars.
To decide whether each particular pair of stars is a
binary system, one needs to compare the parameters
of both objects. In order to do this for different parameters
of all possible pairs in the ensemble is extremely impractical. Therefore, a preliminary list of possible
pairs of stars located closer than 1 pc to each other was
compiled. The determination of the star position in
the three-dimensional space involves the distance to
the Sun defined as $d[pc]=\frac{1000}{\varpi[mas]}$. For compilation
optimization of such a list, which is too time-consuming
for a simple solution of the problem, we implement
the algorithm described in detail in the Appendix.
We completely remove the objects located in the
regions of the sky from our list corresponding to the
known open cluster of Hyades, as well as to the moving
group (according to other sources, the cluster) Mamajek
1, since it is not possible to distinguish binary stars
in these areas. For the pairs from the resulting list
(39 445 pairs), relative motion parameters were calculated—
the difference of proper motions $\Delta\mu$ projection
of the relative motion (in the units of linear velocity $dV = |d|\cdot \Delta \mu$). Figure 1 shows how the distance
between which $d\leq 1$ pc stands out among a subset of
stellar pairs with similar proper motion of the components
in the ensemble of star pairs. Moreover, if the
distance between components is considered, it
becomes noticeable how the ensemble separates into
two subsets also according to its physical separation:
pairs with a large difference of their proper motions
are predominantly wider than those with a small difference
of $\Delta\mu$ (see Fig. 2).

We treat this as a separation between ``random pairings'' and comoving\/binary star pairs, and introduce
an empirical criterion that relates the physical
separation and the difference in proper motions: $\log(\Delta \mu) < 2 - 1.4\sqrt{S}$. Here $S$ is a physical separation in parsec. We include pairs that satisfy this criterion
in our catalog and consider them as candidate binary
systems that are bound (either gravitationally bound
wide binary stars or the members of moving groups).
Independently, the applied criterion can be evaluated
by comparing radial velocities for the star pairs in
which they are known for both components. There are
3593 pairs of such pairs in total (9\%); of these,
1636 satisfy the suggested empirical criterion. \\
For radial velocities, we construct a diagram similar
to Fig. 1 for proper motions (Fig. 3). It is seen that
the radial velocities of the pair components that satisfy
the accepted empirical selection criterion agree well
with each other. Among 46 pairs that satisfy the
accepted criterion, in which the radial velocity difference
between the components exceeds 10 km/s, the
nominal determination errors of the radial velocity in
32 cases are large enough to explain this discrepancy.
A data comparison on the remaining 14 pairs with the
SIMBAD and BDB databases \cite{2015A&C....11..119K} shows that they
include rotating variables (according to the type of
variability in the General Catalog of Variable Stars \cite{2017ARep...61...80S}) and spectroscopic binary stars, which is why
listed Gaia DR2 radial velocity can be associated with, effects other than spatial motion. For two pairs identified
with binary stars HD 53229 and HD 95123, the
radial velocities of the weak components are determined
to be +568 and –715 km/s, respectively, while
the radial velocities of the main components are 14
and 28 km/s. This may be a manifestation of an incorrect
determination effect of radial velocities in dense
stellar fields (See Gaia DR2 Known Issues, as well as \cite{2019MNRAS.486.2618B}), since the angular separation of components in
these pairs is 4.6 and 4.5 mas, respectively.

10 358 pairs of stars that satisfy the above-described
empirical criterion make up a catalog of candidate
wide binary systems with components that have a
common proper motion. The term ``binary star'' is
used for these systems in the text below.

\section{Results}

\begin{figure}[h]
		\includegraphics[width=0.9\linewidth]{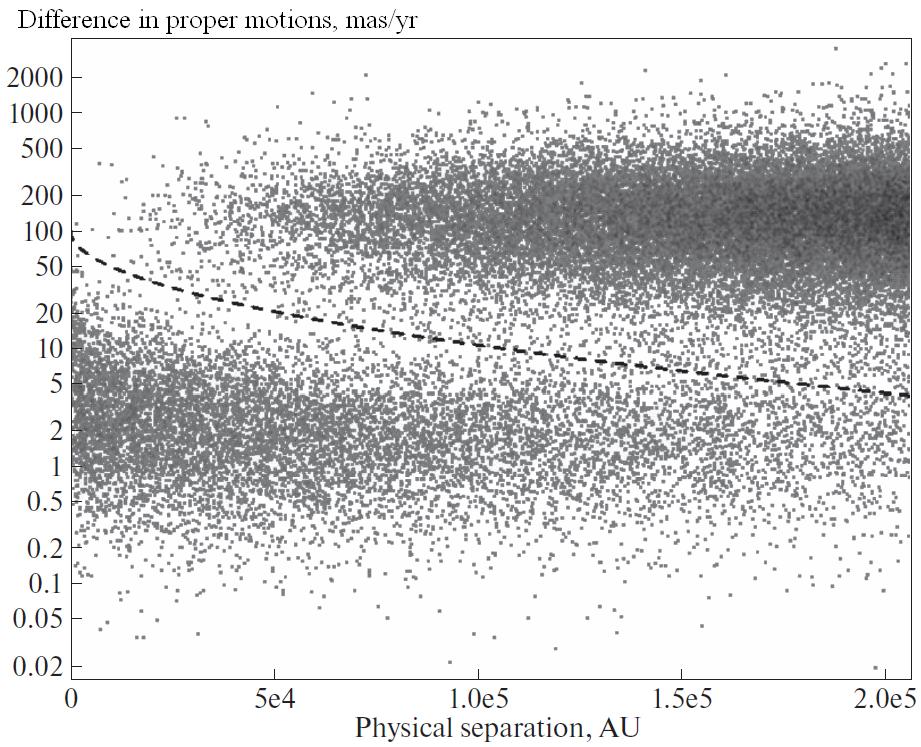}
		\caption{Distribution of stellar pairs in the``physical separation—difference of proper motions'' diagram. Stellar pairs are divided
into two groups, the criterion for the formal separation between them is indicated in the Figure by a dashed line.}
		\label{fig:pm-phsep}
	\end{figure}
	
	\begin{figure}[h]
		\includegraphics[width=0.9\linewidth]{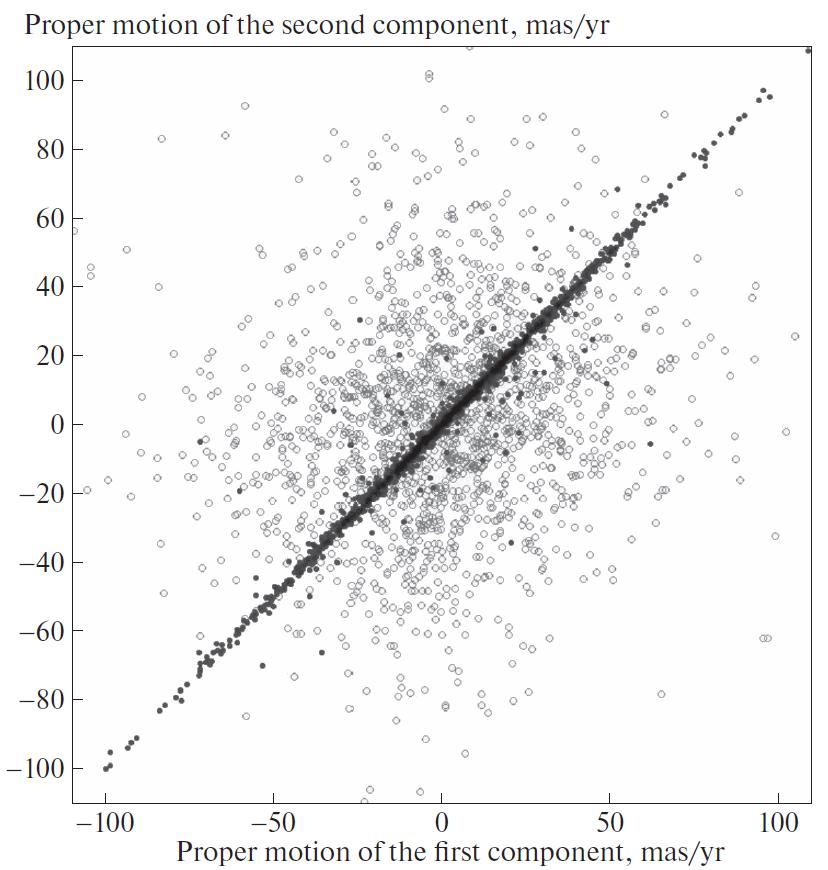}
		\caption{Radial velocity distribution of the components of stellar pairs. The pairs satisfying our limits on proper motion and physical
separation are shown by bold dots, not satisfying limits—by empty dots.}
		\label{fig:rvel}
	\end{figure}
	
	\begin{figure}[h]
		\includegraphics[width=0.9\linewidth]{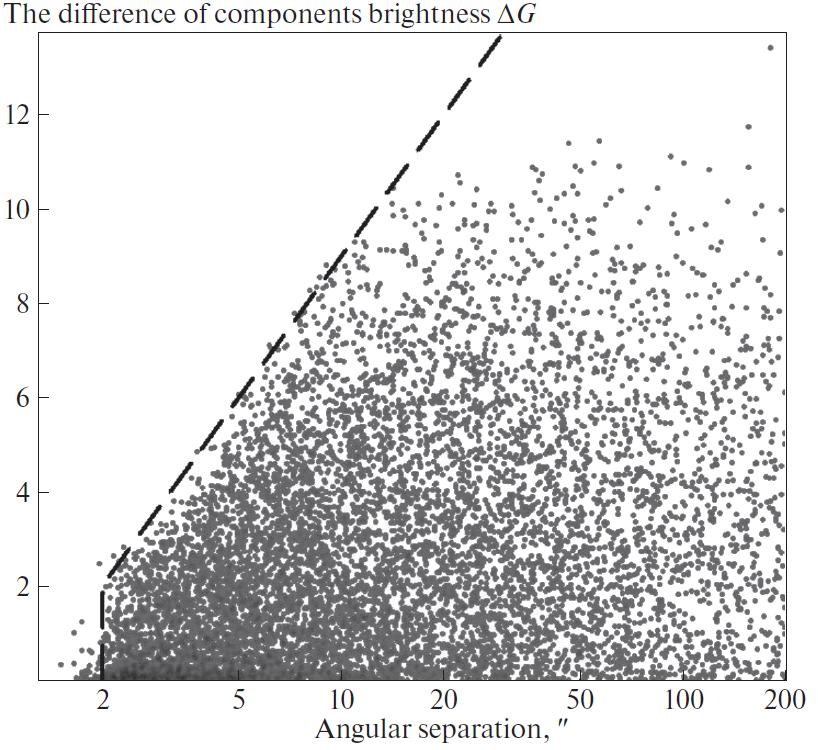}
		\caption{The relationship of the limiting difference of brightness of components and angular separation. The limiting brightness
difference from a certain moment is proportional to the logarithm of the angular separation: by the dashed line is drawn the limit $\Delta G_{lim} \textless 10\cdot\log \rho$  for $\rho \geq 2"$}
		\label{fig:rho_dG}
	\end{figure}

\subsection{Catalog completeness}
Completeness of the created catalog of wide binary
systems at the distance of 10 to 100 pc from the Sun is
determined by several factors. First, it is limited by
completeness of Gaia DR2. It is known that the Gaia
DR2 catalog is substantially incomplete for stars
brighter than approximately $G \approx 6^m$, while it is almost
complete only for the magnitude range $12 \le G \le 17$ \cite{2018A&A...616A...1G}. In addition, the catalog may be incomplete for the
stars with large proper motion (there are more such
stars in the immediate vicinity of the Sun) and in the
dense stellar fields.
Inhomogeneous coverage of the celestial sphere by
Gaia DR2 sources in the microscale (reflecting the
scanning law), as well as the complete removal of the
sources associated with the clusters from the created catalog, should not critically affect the properties of
the obtained sample.
In addition, the completeness of the resulting
binary star catalog is affected by the method of selecting
sources from the Gaia DR2: (i) we selected only
stars with parallaxes; (ii) we applied the filters to select
“astrometrically clean” solutions. In this case, the filter
by the relative parallax error eliminates, predominantly,
more distant stars. The probability of passing
these filters (i.e., having a sufficiently high quality of
astrometric and photometric solutions) particularly is
lower than the average one for unresolved binary stars
and binary components with pronounced orbital
motion, and is equal to zero for pairs with angular distance
between components less than or equal to 2 mas.
For angular distances between components larger than
2 mas, the relation between limiting angular resolution
and the brightness difference between components is
expressed by the relation $\Delta G_{lim} \propto \log \rho$ for $\rho \geq 2"$ (see also a discussion in \cite{ElBadry_Catalogue}). Obviously, this selection
effect predominantly distinguishes pairs with a
small brightness difference at small angular distances (See Fig. 4). \\

Even after excluding the regions of space in which
the stars of the Hyades and Mamajek 1 clusters are
located, the catalog contains a noticeable number of
pairs, with one or both components as members of
more than one pair. As the ratio of the distances
between components in the groups of stars consist
more than of one pair shows, they may be hierarchical
multiple star systems; however, these are mainly stars
of moving groups. About 400 pairs that have a common
component with another pair were found, which
were removed from the catalog in order to avoid distortions
in the characteristic analysis of the ensemble of binary stars. After elimination of multiple stars,
9977 pairs remained in the sample.\\
Generally, one would expect that some of the components
of the catalogued pairs ($10 - 13 \%$, based on statistics on multiplicity from \cite{2014AJ....147...87T, 2013ARA&A..51..269D}) can be represented
by unresolved binary stars (see also the discussion of
pairs with different radial velocities of the components
in Section 2). On the other hand, the catalog compilation
procedure for filtering sources according to the
quality of the astrometric and photometric solutions
should reduce this fraction. A study of solution quality
indicators (RUWE, Flux Excess Factor) performed by
the DR2 authors for known close binary stars ($\rho \le 2$ mas), taken from the identifier catalog of binary and
multiple stars ILB \cite{2016BaltA..25...49M} showed the following:
3537 close pairs were identified as an unresolved
binary star with Gaia DR2 sources for which the parallax
values $\varpi \ge 10$ mas. Out of these sources, Gaia DR2
quality of the solution filter rejected 2180 or $62\%$. For
comparison, only $8\%$ of objects do not pass the same
filter when identifying Gaia DR2 sources with Hipparcos
stars. This suggests that the use of recommended
filters has particularly led to a significant
reduction of the number of components that are unresolved
binary stars.

\subsection{Analysis of sample features}

	\begin{figure}[h]
		\includegraphics[width=0.9\linewidth]{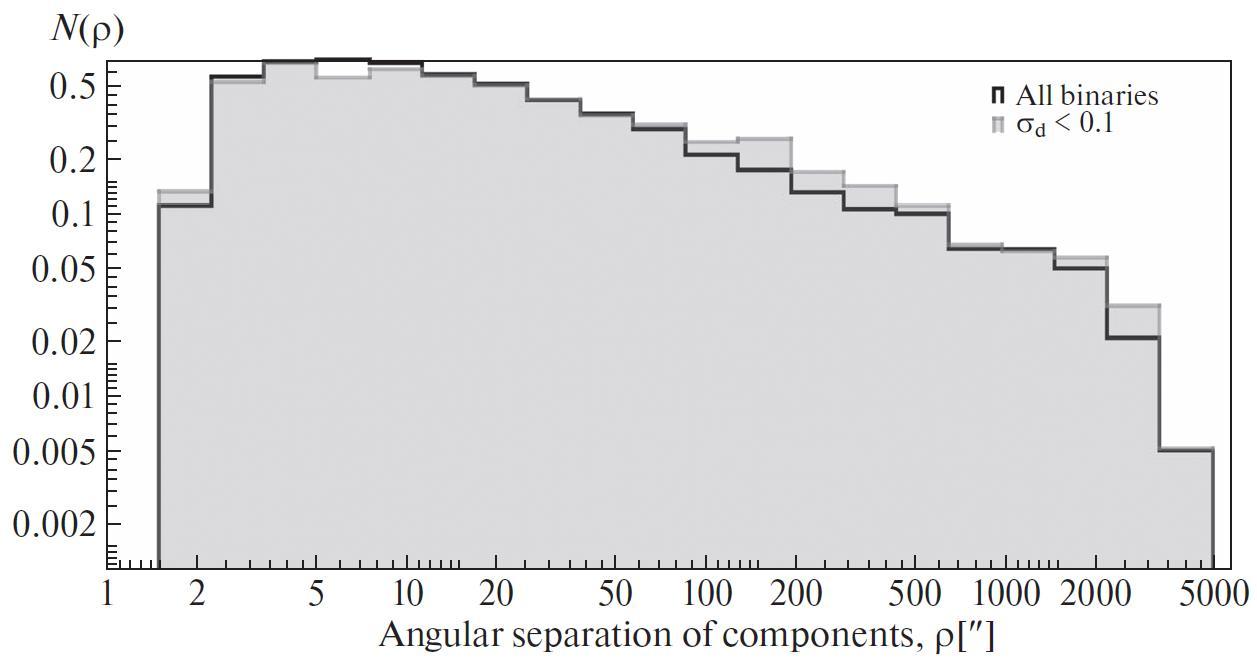}
		\includegraphics[width=0.9\linewidth]{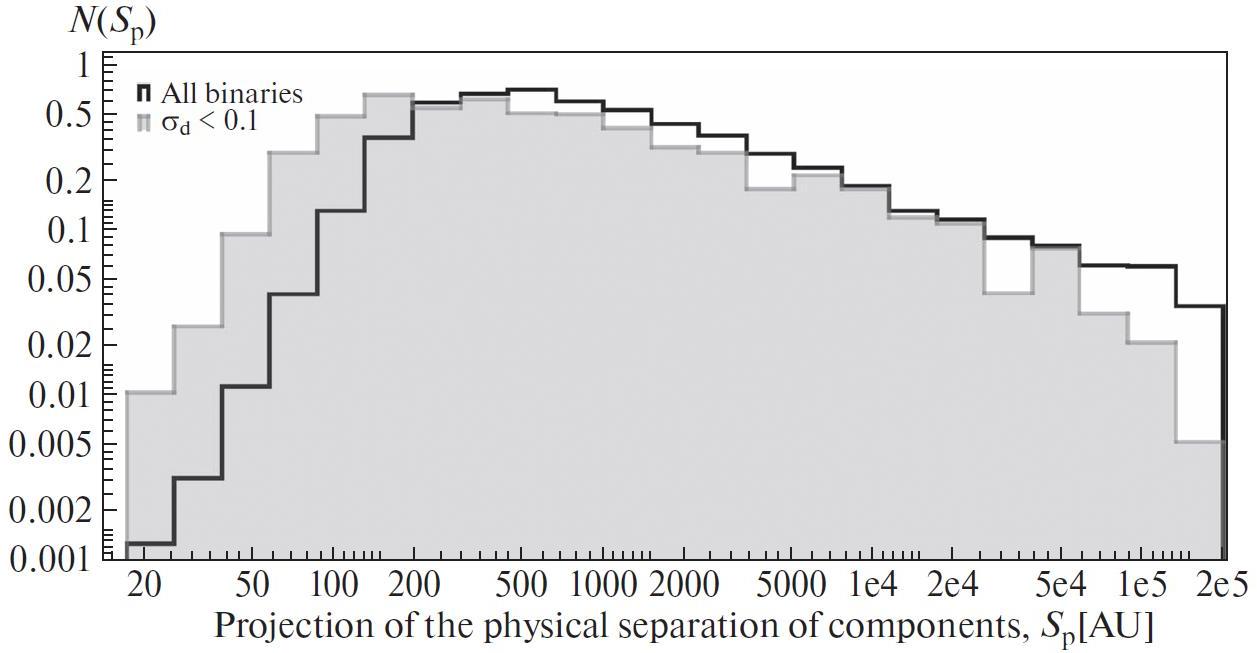}
		\includegraphics[width=0.9\linewidth]{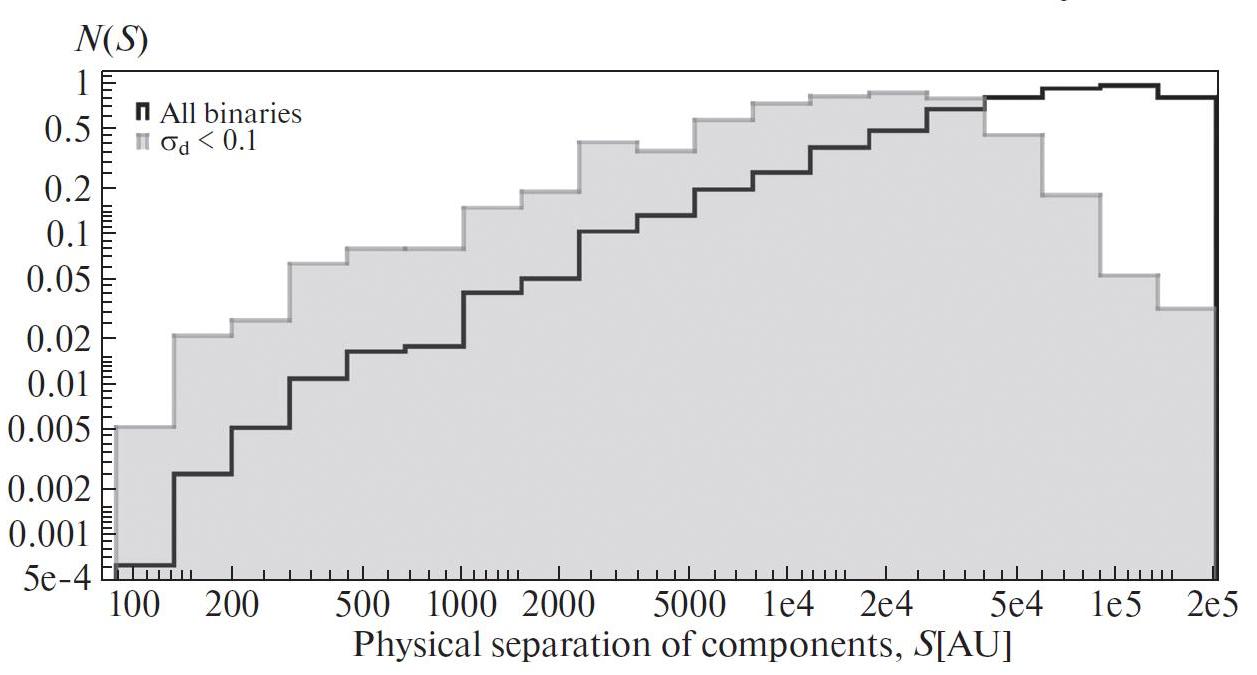}
		\caption{Normalized distributions (top to bottom) over angular separation between components (a), over projection of the physical
separation between components (b), and over the physical separation between components in three-dimensional space (c). The
black line shows all binaries, by gray filling are indicated systems that satisfy requirement of the error in determination of the distance
being not larger than 0.1 pc for both components.}
		\label{fig:phsep}
	\end{figure}
	\begin{figure}[h]
		\includegraphics[width=0.9\linewidth]{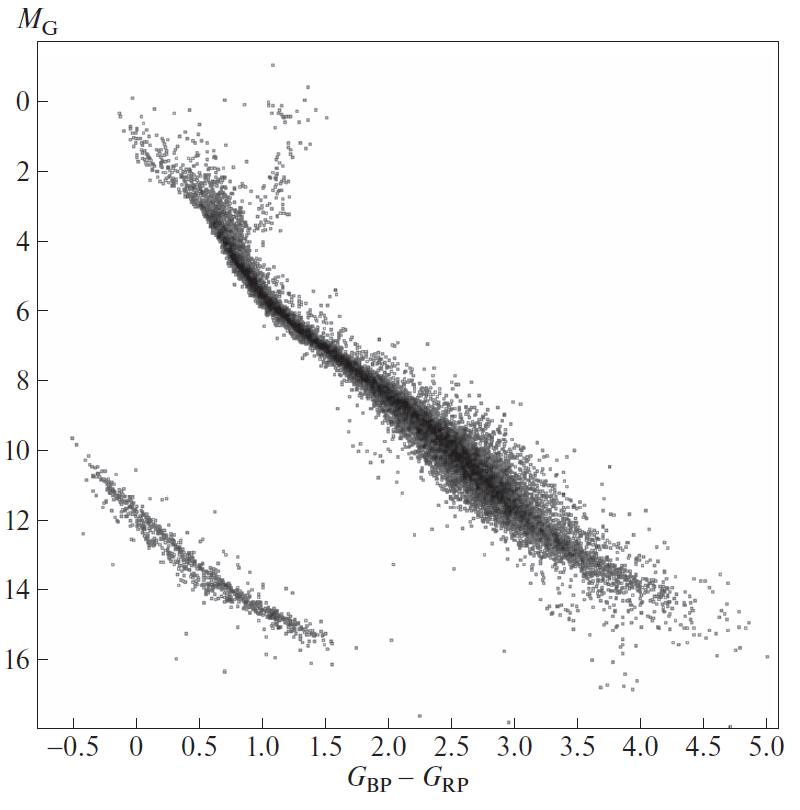}
		\caption{``Color–absolute magnitude'' diagram for all components of the binary stars of the catalog.}
		\label{fig:hr}
	\end{figure}
	
	\begin{figure}[h]
		\includegraphics[width=0.45\linewidth]{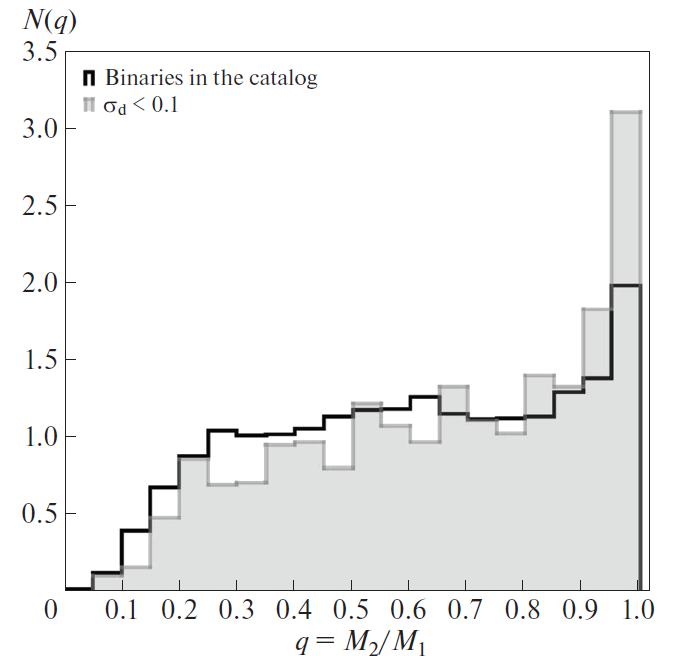}
		\includegraphics[width=0.45\linewidth]{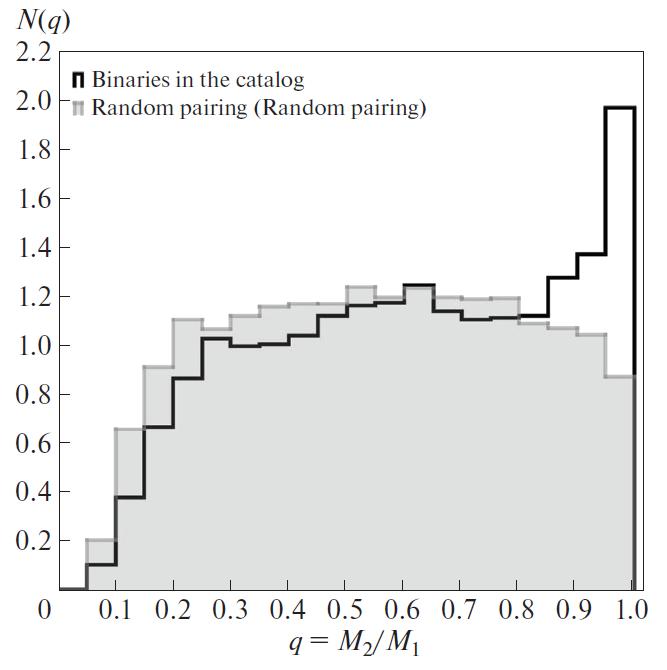}
		\caption{Normalized distributions over mass ratio $ q = \frac{M_1}{M_2}$ Left panel – distributions for the entire catalog (solid line) and for the
subsample with the limit on the maximum error of determination of the distance (filled histogram). Right panel shows comparison
of the distribution for our catalog (solid line) with the distribution for ``random pairing'' — a sample of binary stars obtained
by randomly pairing stars from Gaia (solid line).}
		\label{fig:mass_ratio}
	\end{figure}
	
	\begin{figure}[h]
		\includegraphics[width=0.95\linewidth]{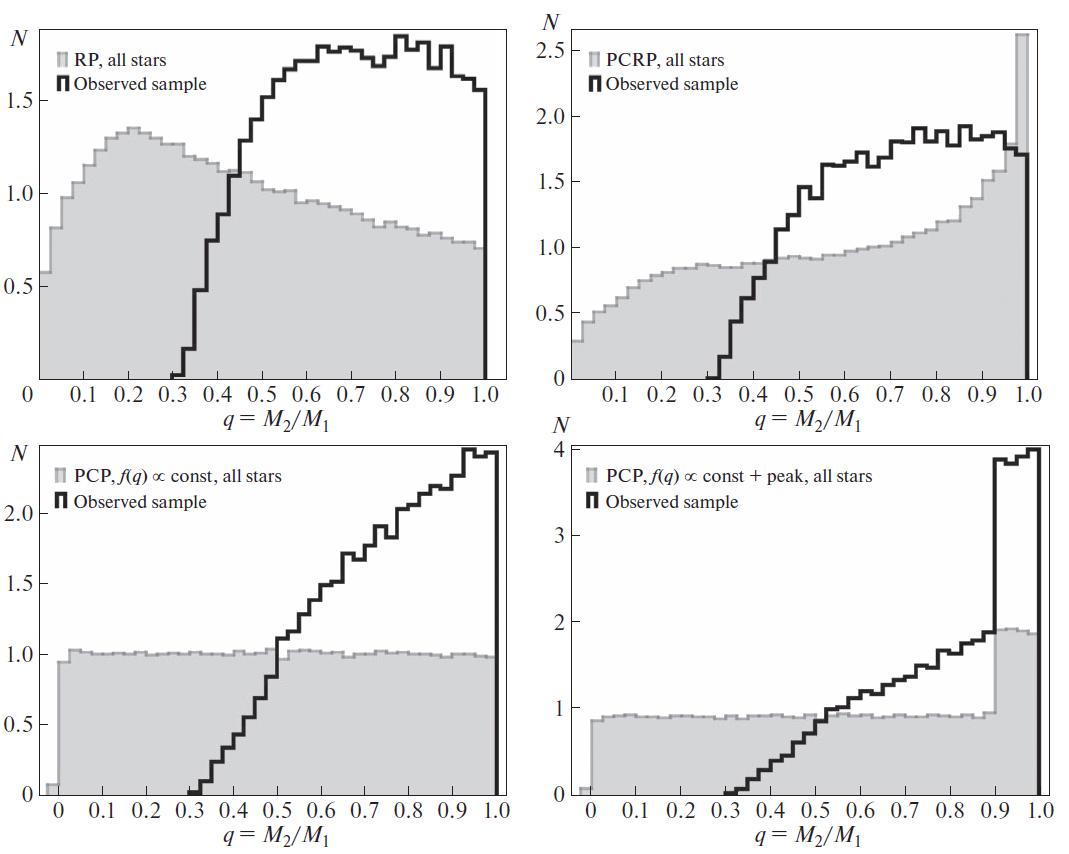}
		\caption{Normalized distribution over mass ratio of components q for a binary star ensemble modeled under various assumptions
about formation of star pairs within 100 pc from the Sun (gray filled histogram) and its observed part with an imitation of the
catalog selectivity function (black contour histogram), see the text for details.}
		\label{fig:mass_ratio_dist}
	\end{figure}
	
	\begin{figure}[h]
		\includegraphics[width=0.95\linewidth]{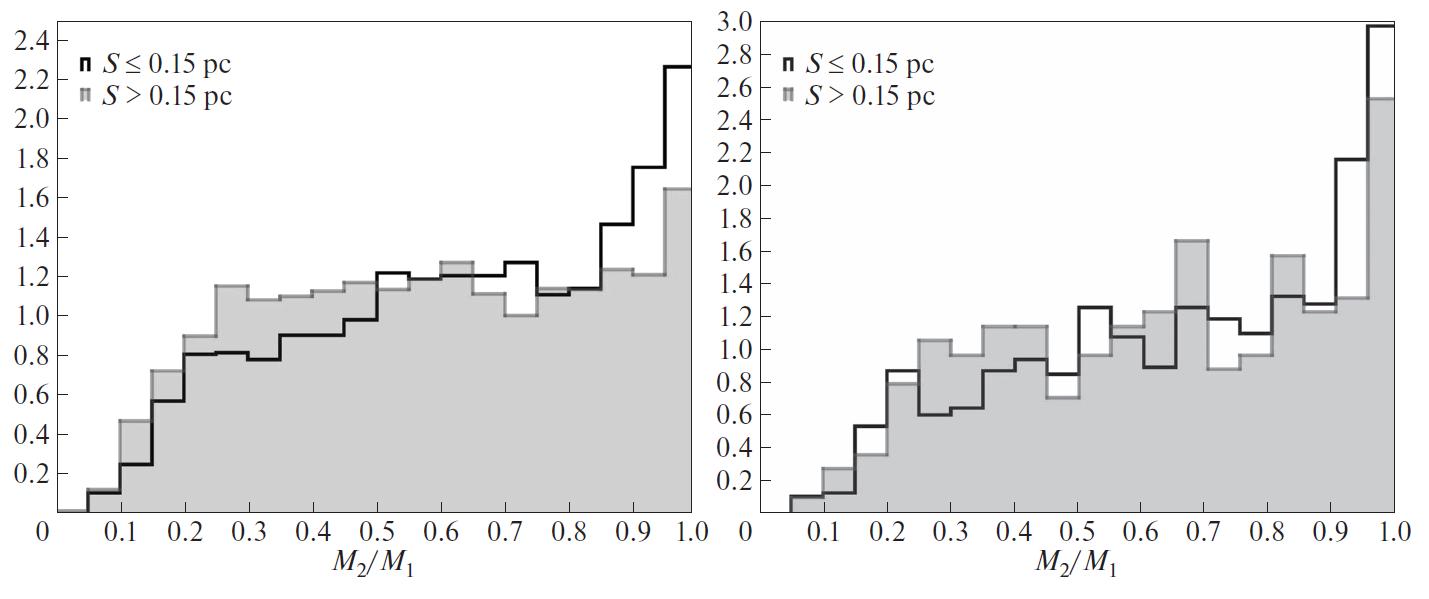}
		\caption{Normalized distributions over mass ratio of components $ q = \frac{M_1}{M_2}$ for binary stars with the separation in the threedimensional
space lower than 0.15 pc (solid line) and exceeding 0.15 pc (filled histogram). Left panel shows distribution for the
entire catalog, right panel – for the subsample with a restriction on the determination error of the distance of 0.1 pc.}
		\label{fig:model}
	\end{figure}

Let us consider the distribution of sample pairs
over the angular separation of components (Fig. 5a), over the distance between components in the projection
onto the celestial sphere (Fig. 5b), and in the linear
space (Fig. 5c). The distribution of the angular
separations between components (Fig. 5a) in the logarithmic
scale turns out to be flat in the $3 \le \rho \le 20 $ mas range and decreases at larger values. Although the parallax value is involved in the calculation of both
linear distances, the accuracy of its determination is
more significant for the distance in the three-dimensional
space. This is reflected in the distributions: the
projection distribution of the distances between components
(Fig. 5b) is similar to the distribution in Fig. 5a, linearly decreasing in the logarithmic scale
beyond 300–400 AU (consistently with the estimate of $10^{2.5}$ A.U. in \cite{ElBadry_Catalogue}).The distribution over the distance
between components in the three-dimensional space
in the logarithmic scale increases to $\approx 0.5$ pc, or $10^5$ A.U. \\

Despite the accepted limitation of the relative parallax
error by $10\%$, for the distances between components
typical for binary stars, small relative parallax
errors can lead to large relative errors in the determined
distances between the stars. To separately consider the
parameter distribution of a more refined, albeit less
complete sample, a subsample of binary stars is introduced
in which the nominal error estimate in the parallax
distance does not exceed 0.1 pc: $\delta_{d} < 0.1 , \delta_{d} = \frac{1000}{\varpi-\delta_{\varpi}} - \frac{1000}{\varpi}$ (henceforth will be called sample I). On figure \ref{fig:phsep} distributions for sample I are represented by a grayfilled
histogram. It can be seen that the introduction of
a limit on the error of the distance significantly
reduces the binary fraction with extremely large separations
(more than 50 000 AU) and leads to a shift of
the distribution maximum in the logarithmic scale to
the lower values. \\
We use absolute magnitudes in the photometric
band G and the color obtained from the difference in
magnitudes in the BP and RP Gaia bands to construct
the Hertzsprung-Russell diagram (Fig. 6). According to the position in the diagram, it is possible to distinguish
the components populating the main-sequence
for which we estimate the masses from absolute magnitudes
in the band G, using Mamajek’s tables\footnote{\url{http://www.pas.rochester.edu/~emamajek/EEM_dwarf_UBVIJHK_colors_Teff.txt}}. Moreover, we neglect the probability that the catalog
contains unresolved binary stars and stars evolving to
the main-sequence, although such objects may be
present in the sample.\\
This allows us to proceed to the mass ratio distribution $q=\frac{mass(B)}{mass(A)}$ for pairs in which both components are
supposedly located on the main-sequence (see Fig. 7).
In the region $0.3 \leq q \leq 0.9$ the distribution looks
close to a planar one. In the region $q \geq 0.95$ an excess
of binary stars is detected (the so-called twin stars,
with components of close mass values). It is important
to find out whether such a feature reflects the nature
of wide binary stars or if it is related to the effects of
sample selection.

\subsection{Catalog selectivity function}

As discussed above, selection effects that affect the
sample are a combination of the Gaia DR2 selectivity
function which has predetermined quality decision filters
for single stars and the pair selectivity function.
We create a sample of random star pairs from Gaia DR2 with $\varpi<15$ mas (to limit the sample size),
selecting them with the same solution quality requirements
that were imposed on the pair components in
Section 2. Using Hertzsprung-Russell diagram, we
select the pairs with both components belonging to the
main sequence, estimate their component masses, and
construct the ratio distribution of the masses of the
weak (“secondary”) and bright (“main”) components.
Figure 7 shows the mass distribution for the
sample of wide binary stars in comparison with a similar
distribution for the synthesized random sample. It
is observed that the selectivity function of Gaia DR2
in combination with the applied filters for the solution
quality of the components cannot be the reason for the
appearance of the “peak of twins”.\\

In order to investigate how the pair selectivity function
associated with the different visibility of stars with
different magnitude contrasts at the same angular distance
can distort the observed distribution of mass
ratios, we will model this effect using binary population
synthesis code \cite{2018NewA...61...24G} for various assumptions about
the mechanisms ``scenario'') of binary system formation \cite{2009A&A...493..979K}. Kroupa \cite{2001MNRAS.322..231K}, initial mass function (IMF)
was used and various scenarios of combining the component
masses were assumed: independent distribution
of them and/or distribution of the component
mass sum according to the IMF. For scenarios assuming
that the distribution of the component mass ratio $f(q)$, is a free parameter, various options were considered,
including $f(q) \propto C$. For the distribution over
semi-major axes of the orbits, a power function $f(a) \propto a^\beta$, was chosen. Since it was found that the
value of the power does not affect the distribution
shape over mass ratio of components, the value $\beta = -1$ was adopted. The distribution over eccentricity of the
orbits was taken to be flat. For each of such combinations
of initial conditions, the vicinity of the Sun
within 100 pc was simulated. Star formation rate over
14 Gyr was assumed to be $15\,e^{-t/\tau}$, where $\tau=7$\ Gyr \cite{2001A&A...365..491N}.\\

Stellar evolution was described by approximate formulas
of Hurley et al. \cite{2000MNRAS.315..543H}. Then, the simulated sample
was compared with the part that would be “visible”
under given conditions, including the requirements
that both components are main-sequence stars, the
restrictions on the apparent magnitude correspond to
the Gaia DR2 restrictions $6^m \le G \le 17^m$, and the pair
selectivity function (Section 3.1). In the studied scenarios,
the catalog selectivity function for binary systems
did not lead to a pronounced peak formation of
close mass components. Moreover, if (in the scenarios
of pair formation allowing an independent distribution
of component mass ratios) a similar peak is introduced
artificially, the adopted “observational” filters do not
significantly affect the presence and shape of this
peak. Figure 8 shows the normalized distributions of
the model sample over mass ratios of components for
the pairs “existing” in the model under given constraints
for some combinations of initial conditions,
compared to the “observable” ones. Presented are the
following scenarios: RP scenario—“random pairing”
(both components are formed independently with masses that are selected from the IMF, top panel, left);
PCRP scenario—“primary constrained random pairing”
(the main component is formed like in RP scenario,
the secondary one – with the mass from the
IMF, provided that it is not more massive than the
main component, top panel, right); and PCP scenario
– “primary constrained pairing” (the main component
is formed like in RP scenario, the mass of the secondary
is determined by an independently specified
distribution over mass ratios of components, bottom
panel for the case of a flat distribution by mass of components
(left), and a flat distribution with a peak (step)
at $q \geq 0.9$ for (right)). In these scenarios, the use of the
catalog selectivity function that limits the pair visibility
with close components with a large brightness difference
does not lead to the formation of a “twins
peak” in the distribution over component mass ratio if
it was absent in the initial sample. The peak in the initial
distribution according to the PCRP scenario disappears
when the observations are simulated, due to
the fact that it is formed by the least massive stars, rejected by the limit of the apparent magnitude. The
peak of the twins in the initial distribution according to
the PCP scenario with a flat distribution and superimposed
peak remains in the observed distribution.

\subsection{Twin stars}

The excess of twin stars among wide binaries
should be recognized as really existing. This result was
obtained independently of a detailed study \cite{ElBadry_Twins} and
supports the conclusions drawn there.\\
The excess of twins becomes less pronounced with
increasing separation of components: Fig. 9 shows
how the distribution over mass ratios becomes flatter
at large physical separations. To the right in the same
Figure, the distribution for sample I with a restriction
on the distance error is shown. In the sample with a
more stringent restriction on the error of the distance
between components, the peak of twins is more narrow
and more pronounced. There can be several reasons
for this, and it is possible that the resulting effect is achieved by their combined effect. First, the average
distance between the components in the sample I is
smaller (and at closer distances, the proportion of systems
with twin components is higher). This may be
due to the fact that the predominant formation mechanism
of twin stars is not effective for the widest pairs.
On the other hand, it can be expected that the introduction
of a restriction on the error in determined distance
for the catalogued stars leads to the decrease in
the contamination of the sample by optical pairs, the
distance between which is underestimated due to the
parallax errors. The admixture of optical pairs in which the distribution of mass ratio falls in the range $0.4 \le q \le 1$ (Fig. \ref{fig:mass_ratio}) should lead to the decrease in the
star fraction with large q. Therefore, the smaller it is,
the more pronounced the ``twin peak'' should be. 
A detailed review and discussion of possible channels
for the formation of an excess of wide binary stars
with close masses of components are presented in \cite{ElBadry_Twins}and references therein. It is important to note that the
most probable mechanism is the competitive accretion
onto the forming pair components in a common protostellar
disk, in which the component with an initially
smaller mass, moving in the orbit with a large radius,
can come up in mass with the initially more massive
star. Currently, the problem of this mechanism is that
the largest observable protostellar disks around forming
binaries have a radius of the order of several hundred
AU (e.g. \cite{2016Natur.538..483T}), while the excess of twin stars,
though decreasing with increasing separation of components,
continues to be significant up to the distances
exceeding $3 \cdot 10^4$ AU.\\

With the further increase of component separation,
the star excess with close masses continues to decrease
and completely disappears for pairs with a calculated separation larger than $8 \cdot 10^4$ AU, for which the distribution
over component mass ratio becomes flat over
the entire range $0.3 \le q \le 1$. Such a mass ratio distribution
can be the result of a combination of a number
of factors (the presence of initially binary stars whose
orbit underwent dynamic broadening \cite{2001ApJ...555..945K}; presence of
a share of pairs formed during cluster decay \cite{2010MNRAS.404.1835K}]; and
admixture of optical pairs). Finally, in the analysis of
the distribution dependence over component mass
ratios on their separation, it should be considered that
the accuracy of the separation estimates itself strongly
depends on the determination accuracy of parallaxes.

\section{Conclusions}
Using Gaia DR2, a catalog of common motion
pairs in the 100 pc vicinity from the Sun was compiled.
Candidate pairs, pre-selected on the base of component
separation in the three-dimensional space, were
filtered by a selected empirical criterion that takes into
account proper motions and physical separation
between components of the pairs, as well as a catalog
including about 10 000 binary and common proper
motion stars. The incompleteness extent of the resulting
catalog is determined mainly by the combination
of the Gaia DR2 catalog selectivity function for single
stars, the Gaia DR2 selectivity function for star pairs,
and the filter system of astrometric and photometric
quality adopted for the selection of Gaia DR2 sources.
In particular, it was shown that there is a relationship
between the limiting brightness difference for a pair of
stars and the angular distance at which they can be
detected, up to 10 mas. The form of this dependence is
suggested and shown that the minimum angular distance
between components in the catalog is 2 mas as the result of the use of solution quality filters, and the
share of unresolved binaries among the catalog pairs is
significantly (possibly up to 60\%) lower than the average
for the field stars. Common motion groups of stars
with the number of members exceeding 2 were
removed from the catalog. It is shown that the distribution
of catalog pairs by separation in the threedimensional
space demonstrates a distribution that
differs from the distribution over the projection of separations
onto celestial sphere due to parallax determination
errors. To reduce this effect, a subsample of
binaries with strong restrictions on the error of determination
of the distance is selected.\\
Using Hertzsprung-Russell diagram, the pairs of
stars are selected, both components of which are presumably
on the main-sequence. For such pairs, the
masses of components are estimated. Independently
from \cite{ElBadry_Twins}], a confirmation of existence of an excess of
binary systems``twins''—with close masses of components
is obtained. It is shown that the detection of
the peak in the catalog sample is not a consequence of
its selectivity function. It is shown that this ``twins
peak'' becomes less pronounced with increasing separation
of components. However, it disappears completely
only at the calculated separations greater than $8 \cdot 10^4$ AU. Like the estimate made in \cite{ElBadry_Twins}, this value
significantly exceeds the size of the observed protostellar
disks, where the binaries are formed. At the same
time, the presence of an excess of twin stars suggests
that they are formed during a process in which the
masses of the components are dependent on each
other. This requires a study of possible mechanisms for
a significant increase of separation of components of
pairs, occurring also during decay of stellar clusters.

\hspace{1cm}
\appendix
\begin{center}
    \begin{large}
    APPENDIX
    \end{large}
\end{center}
\section{The algorithm for compiling the list of stellar pairs}

	\begin{figure}
		\includegraphics[width=1\linewidth]{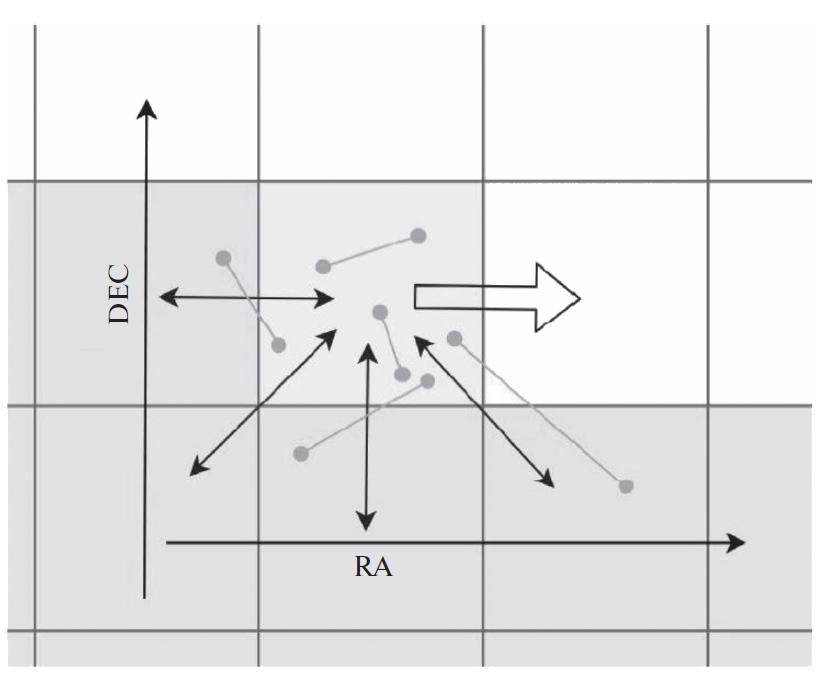}
		\caption{Demonstration of sequential passage through the sections in order to find binaries. The square in the center with diverging
arrows is the area checked at a certain step of the algorithm. Other shaded areas were already covered by the algorithm. Four
arrows directed down and to the left connect the sections the presence of pairs of stars in which will be checked at this step. Gray
circles connected by the lines represent examples of pairs that can be detected at this step, both inside the site and in the ``adjacent''
sites. The arrow to the right indicates which section will be the next, with the larger right ascension. Possible pairs located
in the current section and in sections with the larger declination or larger right ascension will be checked at the later steps of the
algorithm.}
		\label{ManyBoxes}		
	\end{figure}

In this section, we describe the algorithm developed
to create a list of Gaia DR2 source pairs that are
close enough to each other in three-dimensional space
$\alpha, \delta, \omega$. Creation of a list of binaries implies pairwise
operations on the stars, which will take $C N^{2}$of computer
time the multitude of $N$ stars. This will make
such calculations too long for any significantly large
star sample. Of course, it makes sense to compare the
parameters of those stars only that have at least some
chance of being in a bound pair. There is no need to
spend time to be sure that two stars located at a distance
of 50 pc from the Sun in opposite parts of the
celestial sphere are not a physical couple. It is possible
to introduce a criterion that will consider whether it
makes sense to perform detailed calculations for the
given pair of stars (for example, a criterion can be a
large projected angular separation), and to determine
whether this criterion is fulfilled, skipping further calculations
if it is not satisfied. However, this criterion calculation carried out for all pairs of stars in the sample
is still a challenge of complexity $C N^{2}$: backtracking
from further calculations if this criterion is not met,
then $C$ is reduced. \\
It seems more efficient to create an algorithm that
would make a list of potential pairs in the first approximation.
This algorithm can be optimized as much as
possible to reduce the sample in which complexity $C N^{2}$ calculations would be performed. With a list of
candidates for pairs instead of a star list, calculations
can be performed in a row rather than pairwise over all
stars, according to the existing list, which is much
faster. The creation of such algorithm was the first
(and significant) part of this study. Thus, at the first stage of the study, the goal was to
transform the task $C N^2$ into a task $CN$. All stars were
selected as candidates for pairs located in threedimensional
space at a distance less than 1 pc to each
other. The distance was calculated based on the celestial
coordinates of the components $\alpha_i, \delta_i$ and the estimates
of the distances to them, calculated as the reciprocal
of the parallax $\varpi$. Using such estimates is permissible
due to the fact that, in the ensemble under
study, there are sufficiently close stars with characteristic
determination error of the parallax of Gaia DR2 $\epsilon_\varpi/\varpi <10\%$). \\
The next problem was to determine which stars are
subject to pairwise comparison in order to verify that
the proximity criterion is met. Let us look for star pairs
that have a distance from the Sun at least not less than
some fixed value $d$. Obviously, starting with some
angular separation between two stars $\theta$ their physical
separation cannot be less than the assigned allowed
maximum. By solving a simple geometric problem,
this angular separation is equal to $\theta = 2 \cdot \arcsin(p/2d)$ (where $p$ is the maximum physical separation, in our
case—1 pc. Using this, in the sky, we can select rectangular
in coordinates $(\alpha, \delta)$ regions $(\alpha_1,\alpha_2;\delta_1,\delta_2)$ (hereinafter, we will call them ``section''), the width
along both axes will be equal to the above-mentioned
maximum separation. Thus, we would know that a star
at one side of such a section (for example, $\alpha > \alpha_2$)
cannot be a pair to a star on the opposite side ($\alpha < \alpha_1$). It is important to note that later in the Section on compiling
the binary list, the expression ``cannot be a pair''
is used for convenience in the sense ``cannot be closer
than 1 pc from each other''. Partitioning into similar
sections gives us an opportunity to not consider pairs
located at the distance exceeding the width of a
section. It is worth noting that the size of such a rectangular
section over $\alpha$ will be larger than over $\delta$: for
large declinations, the same angular separation corresponds
to a greater distance over right ascension: $|\alpha_2-\alpha_1| = \theta/\cos(\delta)$. \\
Partitioning into sections is done within large areas
equal to 1/4 of a complete circle in right ascension and
about twenty degrees in declination. Within one large
area, the correction ($|\alpha_2-\alpha_1| = \theta/\cos(\delta)$) is accepted
as the largest possible (i.e., it is selected for the largest
declination value within the given area). Thus, each
large area has its own rectangular grid of sections.\\
Upon the creation of the described partition, we
check all possible pairs within one large area and
obtain a gain in the computational time. Thus, for any
star, a pair can be found either in the same region or in
the neighboring one. Thus, two actions are accomplished:
checking the distances between all stars in
pairs within the same section, and checking the distances
between all stars in the neighboring sections
(that is, only the connections between the sections).
The algorithm for performing these two actions is
implemented in such a way that repetitions do not
occur. The execution sequence is graphically depicted
in a Fig. 10 example by one of the steps in the middle
of the passage over a large area. The passage is made in
the direction of increasing right ascension, and is
shifted to the next section of the declinations array,
when the right ascension passage for specific declination
is completed. At each step, a section is selected,
and the following are checked: the presence of star
pairs within the site, the presence of star pairs of this
site, and four out of the eight adjacent sites. The remaining four ``neighborhood'' will be checked in
the next steps (see Fig. 10). \\
Within one large region, the algorithm ensures that
all stars separated by less than 1 pc are added to the list
of potential pairs. After this, the potential pairs formed
by stars located on the opposite sides of the boundaries
between large regions are checked separately. Again,
only those stars that have separation not exceeding the
maximum separation from the boundary of the
regions pass the test. After combining all these regions
for different declinations, only polar regions remain
unchecked, in which the above-described algorithm is
not suitable due to the tendency to infinity of the factor $1/\cos(\delta)$. Verification of the pairs in the circumpolar
regions is implemented for the entire section with a
declination radius selected in such a way that it
exceeds. The final result is a list of pairs for all parts
of the celestial sphere.\\
The entire above-described procedure is performed
for ``layer'', limited by two values of the distance
from the tested stars to the Sun (calculated by
Gaia parallax): $d_1 < d < d_2$. Respectively, the maximum
angular separation $\theta$is determined by the smallest
possible distance $d_1$.For this, the $10 \leq d \leq 100$ pc
sample was divided into four layers: from 10 to 25 pc,
from 25 to 50 pc, from 50 to 75 pc, and from 75 to
100 pc. For each layer, the values of $\theta$ and partition
into sections are determined separately. Thus, despite
the larger number of stars in the outer layers, we partially compensate for the increase in computational
time by reducing the size of the sections. The pairs
formed by stars located on the opposite sides of the
boundaries of these spherical layers are checked separately:
this is achieved by taking these layers with an
overlap equal to the maximum selected physical separation
(in our case, 1 pc) and removal of the pairs,
both of which are in the overlapping region of one of
these sets. Thus, we avoid including them in the list of
pairs more than once. Due to the overlap, those pairs
that are located on the opposite sides of the layer
boundary are taken into account.
\section{Acknowledgments}
This work has made use of data from the European
Space Agency (ESA) mission {\it Gaia}  processed by the Gaia
Data Processing and Analysis Consortium (DPAC). Funding
for the DPAC has been provided by national institutions,
in particular the institutions participating in the Gaia
Multilateral Agreement. The data from SIMBAD database
supported by CDS (Strasbourg, France) were used. In
selection of stellar pairs, astropy package \cite{astropy:2013}, \cite{astropy:2018} was used. Processing and analysis of the data were made using Topcat\footnote{http://www.starlink.ac.uk/topcat/}. The authors acknowledge A.V. Tutukov and D. Chulkov for useful discussions.

The study was partially supported by Russian Foundation
for Basic Research (Project no. 19-07-01198).
\printbibliography

@ARTICLE{binariesDR2feb,
	author = {{Jim{\'e}nez-Esteban}, F.~M. and {Solano}, E. and {Rodrigo}, C.
	},
	title = "{A Catalog of Wide Binary and Multiple Systems of Bright Stars from Gaia-DR2 and the Virtual Observatory}",
	journal = {\aj},
	archivePrefix = "arXiv",
	eprint = {1901.03730},
	primaryClass = "astro-ph.SR",
	year = 2019,
	month = feb,
	volume = 157,
	eid = {78},
	pages = {78},
	doi = {10.3847/1538-3881/aafacc},
	adsurl = {http://adsabs.harvard.edu/abs/2019AJ....157...78J}
}

@ARTICLE{WideBinariesKepler,
	author = {{Godoy-Rivera}, D. and {Chanam{\'e}}, J.},
	title = "{On the identification of wide binaries in the Kepler field}",
	journal = {\mnras},
	archivePrefix = "arXiv",
	eprint = {1807.00009},
	primaryClass = "astro-ph.SR",
	year = 2018,
	month = oct,
	volume = 479,
	pages = {4440-4469},
	doi = {10.1093/mnras/sty1736},
	adsurl = {http://adsabs.harvard.edu/abs/2018MNRAS.479.4440G}
}

@ARTICLE{GDR2_AstrometricSolution,
	author = {{Lindegren}, L. and {Hern{\'a}ndez}, J. and {Bombrun}, A. and 
	{Klioner}, S. and {Bastian}, U. and {Ramos-Lerate}, M. and {de Torres}, A. and 
	{Steidelm{\"u}ller}, H. and {Stephenson}, C. and {Hobbs}, D. and 
	{Lammers}, U. and {Biermann}, M. and {Geyer}, R. and {Hilger}, T. and 
	{Michalik}, D. and {Stampa}, U. and {McMillan}, P.~J. and {Casta{\~n}eda}, J. and 
	{Clotet}, M. and {Comoretto}, G. and {Davidson}, M. and {Fabricius}, C. and 
	{Gracia}, G. and {Hambly}, N.~C. and {Hutton}, A. and {Mora}, A. and 
	{Portell}, J. and {van Leeuwen}, F. and {Abbas}, U. and {Abreu}, A. and 
	{Altmann}, M. and {Andrei}, A. and {Anglada}, E. and {Balaguer-N{\'u}{\~n}ez}, L. and 
	{Barache}, C. and {Becciani}, U. and {Bertone}, S. and {Bianchi}, L. and 
	{Bouquillon}, S. and {Bourda}, G. and {Br{\"u}semeister}, T. and 
	{Bucciarelli}, B. and {Busonero}, D. and {Buzzi}, R. and {Cancelliere}, R. and 
	{Carlucci}, T. and {Charlot}, P. and {Cheek}, N. and {Crosta}, M. and 
	{Crowley}, C. and {de Bruijne}, J. and {de Felice}, F. and {Drimmel}, R. and 
	{Esquej}, P. and {Fienga}, A. and {Fraile}, E. and {Gai}, M. and 
	{Garralda}, N. and {Gonz{\'a}lez-Vidal}, J.~J. and {Guerra}, R. and 
	{Hauser}, M. and {Hofmann}, W. and {Holl}, B. and {Jordan}, S. and 
	{Lattanzi}, M.~G. and {Lenhardt}, H. and {Liao}, S. and {Licata}, E. and 
	{Lister}, T. and {L{\"o}ffler}, W. and {Marchant}, J. and {Martin-Fleitas}, J.-M. and 
	{Messineo}, R. and {Mignard}, F. and {Morbidelli}, R. and {Poggio}, E. and 
	{Riva}, A. and {Rowell}, N. and {Salguero}, E. and {Sarasso}, M. and 
	{Sciacca}, E. and {Siddiqui}, H. and {Smart}, R.~L. and {Spagna}, A. and 
	{Steele}, I. and {Taris}, F. and {Torra}, J. and {van Elteren}, A. and 
	{van Reeven}, W. and {Vecchiato}, A.},
	title = "{Gaia Data Release 2. The astrometric solution}",
	journal = {\aap},
	archivePrefix = "arXiv",
	eprint = {1804.09366},
	primaryClass = "astro-ph.IM",
	year = 2018,
	month = aug,
	volume = 616,
	eid = {A2},
	pages = {A2},
	doi = {10.1051/0004-6361/201832727},
	adsurl = {http://adsabs.harvard.edu/abs/2018A%26A...616A...2L}
}

@ARTICLE{ElBadry_Catalogue,
	author = {{El-Badry}, Kareem and {Rix}, Hans-Walter},
	title = "{Imprints of white dwarf recoil in the separation distribution of Gaia wide binaries}",
	journal = {\mnras},
	year = "2018",
	month = "Nov",
	volume = {480},
	number = {4},
	pages = {4884-4902},
	doi = {10.1093/mnras/sty2186},
	archivePrefix = {arXiv},
	eprint = {1807.06011},
	primaryClass = {astro-ph.SR},
	adsurl = {https://ui.adsabs.harvard.edu/abs/2018MNRAS.480.4884E}
}

@ARTICLE{ElBadry_Twins,
	author = {{El-Badry}, Kareem and {Rix}, Hans-Walter and {Tian}, Haijun and
	{Duch{\^e}ne}, Gaspard and {Moe}, Maxwell},
	title = "{Discovery of an equal-mass `twin' binary population reaching 1000 + au separations}",
	journal = {\mnras},
	year = "2019",
	month = "Nov",
	volume = {489},
	number = {4},
	pages = {5822-5857},
	doi = {10.1093/mnras/stz2480},
	archivePrefix = {arXiv},
	eprint = {1906.10128},
	primaryClass = {astro-ph.SR},
	adsurl = {https://ui.adsabs.harvard.edu/abs/2019MNRAS.489.5822E}
}

@ARTICLE{Catalogue,
       author = {{Sapozhnikov}, S.~A. and {Kovaleva}, D.~A. and {Malkov}, O.~Y.},
        title = "{Catalogue of Visual Binaries in Gaia DR2}",
      journal = {INASAN Science Reports},
         year = 2019,
        month = oct,
       volume = {3},
        pages = {366-371},
          doi = {10.26087/INASAN.2019.3.1.057},
       adsurl = {https://ui.adsabs.harvard.edu/abs/2019INASR...3..366S}
}

@ARTICLE{2015PASP..127..994B,
	author = {{Bailer-Jones}, Coryn A.~L.},
	title = "{Estimating Distances from Parallaxes}",
	journal = {\pasp},
	year = "2015",
	month = "Oct",
	volume = {127},
	number = {956},
	pages = {994},
	doi = {10.1086/683116},
	archivePrefix = {arXiv},
	eprint = {1507.02105},
	primaryClass = {astro-ph.IM},
	adsurl = {https://ui.adsabs.harvard.edu/abs/2015PASP..127..994B}
}

@ARTICLE{2017AstBu..72..122R,
	author = {{Rastorguev}, A.~S. and {Utkin}, N.~D. and {Zabolotskikh}, M.~V. and
	{Dambis}, A.~K. and {Bajkova}, A.~T. and {Bobylev}, V.~V.},
	title = "{Galactic masers: Kinematics, spiral structure and the disk dynamic state}",
	journal = {Astrophysical Bulletin},
	year = "2017",
	month = "Apr",
	volume = {72},
	number = {2},
	pages = {122-140},
	doi = {10.1134/S1990341317020043},
	archivePrefix = {arXiv},
	eprint = {1603.09124},
	primaryClass = {astro-ph.GA},
	adsurl = {https://ui.adsabs.harvard.edu/abs/2017AstBu..72..122R}
}

@ARTICLE{2018A&A...616A...1G,
	author = {{Gaia Collaboration} and {Brown}, A.~G.~A. and {Vallenari}, A. and
	{Prusti}, T. and {de Bruijne}, J.~H.~J. and {Babusiaux}, C. and
	{Bailer-Jones}, C.~A.~L. and {Biermann}, M. and {Evans}, D.~W. and
	{Eyer}, L. and {Jansen}, F. and {Jordi}, C. and {Klioner}, S.~A. and
	{Lammers}, U. and {Lindegren}, L. and {Luri}, X. and {Mignard}, F. and
	{Panem}, C. and {Pourbaix}, D. and {Randich}, S. and {Sartoretti}, P. and
	{Siddiqui}, H.~I. and {Soubiran}, C. and {van Leeuwen}, F. and
	{Walton}, N.~A. and {Arenou}, F. and {Bastian}, U. and {Cropper}, M. and
	{Drimmel}, R. and {Katz}, D. and {Lattanzi}, M.~G. and {Bakker}, J. and
	{Cacciari}, C. and {Casta{\~n}eda}, J. and {Chaoul}, L. and
	{Cheek}, N. and {De Angeli}, F. and {Fabricius}, C. and {Guerra}, R. and
	{Holl}, B. and {Masana}, E. and {Messineo}, R. and {Mowlavi}, N. and
	{Nienartowicz}, K. and {Panuzzo}, P. and {Portell}, J. and
	{Riello}, M. and {Seabroke}, G.~M. and {Tanga}, P. and
	{Th{\'e}venin}, F. and {Gracia-Abril}, G. and {Comoretto}, G. and
	{Garcia-Reinaldos}, M. and {Teyssier}, D. and {Altmann}, M. and
	{Andrae}, R. and {Audard}, M. and {Bellas-Velidis}, I. and
	{Benson}, K. and {Berthier}, J. and {Blomme}, R. and {Burgess}, P. and
	{Busso}, G. and {Carry}, B. and {Cellino}, A. and {Clementini}, G. and
	{Clotet}, M. and {Creevey}, O. and {Davidson}, M. and {De Ridder}, J. and
	{Delchambre}, L. and {Dell'Oro}, A. and {Ducourant}, C. and
	{Fern{\'a}ndez-Hern{\'a}ndez}, J. and {Fouesneau}, M. and
	{Fr{\'e}mat}, Y. and {Galluccio}, L. and {Garc{\'\i}a-Torres}, M. and
	{Gonz{\'a}lez-N{\'u}{\~n}ez}, J. and {Gonz{\'a}lez-Vidal}, J.~J. and
	{Gosset}, E. and {Guy}, L.~P. and {Halbwachs}, J. -L. and
	{Hambly}, N.~C. and {Harrison}, D.~L. and {Hern{\'a}ndez}, J. and
	{Hestroffer}, D. and {Hodgkin}, S.~T. and {Hutton}, A. and
	{Jasniewicz}, G. and {Jean-Antoine-Piccolo}, A. and {Jordan}, S. and
	{Korn}, A.~J. and {Krone-Martins}, A. and {Lanzafame}, A.~C. and
	{Lebzelter}, T. and {L{\"o}ffler}, W. and {Manteiga}, M. and
	{Marrese}, P.~M. and {Mart{\'\i}n-Fleitas}, J.~M. and {Moitinho}, A. and                   9
	{Mora}, A. and {Muinonen}, K. and {Osinde}, J. and {Pancino}, E. and
	{Pauwels}, T. and {Petit}, J. -M. and {Recio-Blanco}, A. and
	{Richards}, P.~J. and {Rimoldini}, L. and {Robin}, A.~C. and
	{Sarro}, L.~M. and {Siopis}, C. and {Smith}, M. and {Sozzetti}, A. and
	{S{\"u}veges}, M. and {Torra}, J. and {van Reeven}, W. and {Abbas}, U. and
	{Abreu Aramburu}, A. and {Accart}, S. and {Aerts}, C. and
	{Altavilla}, G. and {{\'A}lvarez}, M.~A. and {Alvarez}, R. and
	{Alves}, J. and {Anderson}, R.~I. and {Andrei}, A.~H. and
	{Anglada Varela}, E. and {Antiche}, E. and {Antoja}, T. and
	{Arcay}, B. and {Astraatmadja}, T.~L. and {Bach}, N. and
	{Baker}, S.~G. and {Balaguer-N{\'u}{\~n}ez}, L. and {Balm}, P. and
	{Barache}, C. and {Barata}, C. and {Barbato}, D. and {Barblan}, F. and
	{Barklem}, P.~S. and {Barrado}, D. and {Barros}, M. and
	{Barstow}, M.~A. and {Bartholom{\'e} Mu{\~n}oz}, S. and
	{Bassilana}, J. -L. and {Becciani}, U. and {Bellazzini}, M. and
	{Berihuete}, A. and {Bertone}, S. and {Bianchi}, L. and
	{Bienaym{\'e}}, O. and {Blanco-Cuaresma}, S. and {Boch}, T. and
	{Boeche}, C. and {Bombrun}, A. and {Borrachero}, R. and {Bossini}, D. and
	{Bouquillon}, S. and {Bourda}, G. and {Bragaglia}, A. and
	{Bramante}, L. and {Breddels}, M.~A. and {Bressan}, A. and
	{Brouillet}, N. and {Br{\"u}semeister}, T. and {Brugaletta}, E. and
	{Bucciarelli}, B. and {Burlacu}, A. and {Busonero}, D. and
	{Butkevich}, A.~G. and {Buzzi}, R. and {Caffau}, E. and
	{Cancelliere}, R. and {Cannizzaro}, G. and {Cantat-Gaudin}, T. and
	{Carballo}, R. and {Carlucci}, T. and {Carrasco}, J.~M. and
	{Casamiquela}, L. and {Castellani}, M. and {Castro-Ginard}, A. and
	{Charlot}, P. and {Chemin}, L. and {Chiavassa}, A. and {Cocozza}, G. and
	{Costigan}, G. and {Cowell}, S. and {Crifo}, F. and {Crosta}, M. and
	{Crowley}, C. and {Cuypers}, J. and {Dafonte}, C. and {Damerdji}, Y. and
	{Dapergolas}, A. and {David}, P. and {David}, M. and {de Laverny}, P. and
	{De Luise}, F. and {De March}, R. and {de Martino}, D. and
	{de Souza}, R. and {de Torres}, A. and {Debosscher}, J. and
	{del Pozo}, E. and {Delbo}, M. and {Delgado}, A. and {Delgado}, H.~E. and
	{Di Matteo}, P. and {Diakite}, S. and {Diener}, C. and {Distefano}, E. and
	{Dolding}, C. and {Drazinos}, P. and {Dur{\'a}n}, J. and
	{Edvardsson}, B. and {Enke}, H. and {Eriksson}, K. and {Esquej}, P. and
	{Eynard Bontemps}, G. and {Fabre}, C. and {Fabrizio}, M. and
	{Faigler}, S. and {Falc{\~a}o}, A.~J. and {Farr{\`a}s Casas}, M. and
	{Federici}, L. and {Fedorets}, G. and {Fernique}, P. and
	{Figueras}, F. and {Filippi}, F. and {Findeisen}, K. and {Fonti}, A. and
	{Fraile}, E. and {Fraser}, M. and {Fr{\'e}zouls}, B. and {Gai}, M. and
	{Galleti}, S. and {Garabato}, D. and {Garc{\'\i}a-Sedano}, F. and
	{Garofalo}, A. and {Garralda}, N. and {Gavel}, A. and {Gavras}, P. and
	{Gerssen}, J. and {Geyer}, R. and {Giacobbe}, P. and {Gilmore}, G. and
	{Girona}, S. and {Giuffrida}, G. and {Glass}, F. and {Gomes}, M. and
	{Granvik}, M. and {Gueguen}, A. and {Guerrier}, A. and {Guiraud}, J. and
	{Guti{\'e}rrez-S{\'a}nchez}, R. and {Haigron}, R. and
	{Hatzidimitriou}, D. and {Hauser}, M. and {Haywood}, M. and
	{Heiter}, U. and {Helmi}, A. and {Heu}, J. and {Hilger}, T. and
	{Hobbs}, D. and {Hofmann}, W. and {Holland}, G. and {Huckle}, H.~E. and
	{Hypki}, A. and {Icardi}, V. and {Jan{\ss}en}, K. and
	{Jevardat de Fombelle}, G. and {Jonker}, P.~G. and
	{Juh{\'a}sz}, {\'A}. L. and {Julbe}, F. and {Karampelas}, A. and
	{Kewley}, A. and {Klar}, J. and {Kochoska}, A. and {Kohley}, R. and
	{Kolenberg}, K. and {Kontizas}, M. and {Kontizas}, E. and
	{Koposov}, S.~E. and {Kordopatis}, G. and {Kostrzewa-Rutkowska}, Z. and
	{Koubsky}, P. and {Lambert}, S. and {Lanza}, A.~F. and {Lasne}, Y. and
	{Lavigne}, J. -B. and {Le Fustec}, Y. and {Le Poncin-Lafitte}, C. and
	{Lebreton}, Y. and {Leccia}, S. and {Leclerc}, N. and
	{Lecoeur-Taibi}, I. and {Lenhardt}, H. and {Leroux}, F. and {Liao}, S. and
	{Licata}, E. and {Lindstr{\o}m}, H.~E.~P. and {Lister}, T.~A. and
	{Livanou}, E. and {Lobel}, A. and {L{\'o}pez}, M. and {Managau}, S. and
	{Mann}, R.~G. and {Mantelet}, G. and {Marchal}, O. and
	{Marchant}, J.~M. and {Marconi}, M. and {Marinoni}, S. and
	{Marschalk{\'o}}, G. and {Marshall}, D.~J. and {Martino}, M. and
	{Marton}, G. and {Mary}, N. and {Massari}, D. and
	{Matijevi{\v{c}}}, G. and {Mazeh}, T. and {McMillan}, P.~J. and
	{Messina}, S. and {Michalik}, D. and {Millar}, N.~R. and {Molina}, D. and
	{Molinaro}, R. and {Moln{\'a}r}, L. and {Montegriffo}, P. and
	{Mor}, R. and {Morbidelli}, R. and {Morel}, T. and {Morris}, D. and
	{Mulone}, A.~F. and {Muraveva}, T. and {Musella}, I. and
	{Nelemans}, G. and {Nicastro}, L. and {Noval}, L. and {O'Mullane}, W. and
	{Ord{\'e}novic}, C. and {Ord{\'o}{\~n}ez-Blanco}, D. and {Osborne}, P. and
	{Pagani}, C. and {Pagano}, I. and {Pailler}, F. and {Palacin}, H. and
	{Palaversa}, L. and {Panahi}, A. and {Pawlak}, M. and
	{Piersimoni}, A.~M. and {Pineau}, F. -X. and {Plachy}, E. and
	{Plum}, G. and {Poggio}, E. and {Poujoulet}, E. and {Pr{\v{s}}a}, A. and
	{Pulone}, L. and {Racero}, E. and {Ragaini}, S. and {Rambaux}, N. and
	{Ramos-Lerate}, M. and {Regibo}, S. and {Reyl{\'e}}, C. and
	{Riclet}, F. and {Ripepi}, V. and {Riva}, A. and {Rivard}, A. and
	{Rixon}, G. and {Roegiers}, T. and {Roelens}, M. and
	{Romero-G{\'o}mez}, M. and {Rowell}, N. and {Royer}, F. and
	{Ruiz-Dern}, L. and {Sadowski}, G. and {Sagrist{\`a} Sell{\'e}s}, T. and
	{Sahlmann}, J. and {Salgado}, J. and {Salguero}, E. and {Sanna}, N. and
	{Santana-Ros}, T. and {Sarasso}, M. and {Savietto}, H. and
	{Schultheis}, M. and {Sciacca}, E. and {Segol}, M. and
	{Segovia}, J.~C. and {S{\'e}gransan}, D. and {Shih}, I. -C. and
	{Siltala}, L. and {Silva}, A.~F. and {Smart}, R.~L. and {Smith}, K.~W. and
	{Solano}, E. and {Solitro}, F. and {Sordo}, R. and {Soria Nieto}, S. and
	{Souchay}, J. and {Spagna}, A. and {Spoto}, F. and {Stampa}, U. and
	{Steele}, I.~A. and {Steidelm{\"u}ller}, H. and {Stephenson}, C.~A. and
	{Stoev}, H. and {Suess}, F.~F. and {Surdej}, J. and {Szabados}, L. and
	{Szegedi-Elek}, E. and {Tapiador}, D. and {Taris}, F. and {Tauran}, G. and
	{Taylor}, M.~B. and {Teixeira}, R. and {Terrett}, D. and {Teyssand
	ier}, P. and {Thuillot}, W. and {Titarenko}, A. and {Torra Clotet}, F. and
	{Turon}, C. and {Ulla}, A. and {Utrilla}, E. and {Uzzi}, S. and
	{Vaillant}, M. and {Valentini}, G. and {Valette}, V. and
	{van Elteren}, A. and {Van Hemelryck}, E. and {van Leeuwen}, M. and
	{Vaschetto}, M. and {Vecchiato}, A. and {Veljanoski}, J. and
	{Viala}, Y. and {Vicente}, D. and {Vogt}, S. and {von Essen}, C. and
	{Voss}, H. and {Votruba}, V. and {Voutsinas}, S. and {Walmsley}, G. and
	{Weiler}, M. and {Wertz}, O. and {Wevers}, T. and {Wyrzykowski}, {\L}. and
	{Yoldas}, A. and {{\v{Z}}erjal}, M. and {Ziaeepour}, H. and
	{Zorec}, J. and {Zschocke}, S. and {Zucker}, S. and {Zurbach}, C. and
	{Zwitter}, T.},
	title = "{Gaia Data Release 2. Summary of the contents and survey properties}",
	journal = {\aap},
	year = "2018",
	month = "Aug",
	volume = {616},
	eid = {A1},
	pages = {A1},
	doi = {10.1051/0004-6361/201833051},
	archivePrefix = {arXiv},
	eprint = {1804.09365},
	primaryClass = {astro-ph.GA},
	adsurl = {https://ui.adsabs.harvard.edu/abs/2018A&A...616A...1G}
}

@ARTICLE{2019MNRAS.486.2618B,
	author = {{Boubert}, D. and {Strader}, J. and {Aguado}, D. and {Seabroke}, G. and
	{Koposov}, S.~E. and {Sanders}, J.~L. and {Swihart}, S. and
	{Chomiuk}, L. and {Evans}, N.~W.},
	title = "{Lessons from the curious case of the `fastest' star in Gaia DR2}",
	journal = {\mnras},
	year = "2019",
	month = "Jun",
	volume = {486},
	number = {2},
	pages = {2618-2630},
	doi = {10.1093/mnras/stz253},
	archivePrefix = {arXiv},
	eprint = {1901.10460},
	primaryClass = {astro-ph.SR},
	adsurl = {https://ui.adsabs.harvard.edu/abs/2019MNRAS.486.2618B}
}

@ARTICLE{2018NewA...61...24G,
	author = {{Gebrehiwot}, Y.~M. and {Tessema}, S.~B. and {Malkov}, O. Yu. and
	{Kovaleva}, D.~A. and {Sytov}, A. Yu. and {Tutukov}, A.~V.},
	title = "{Star formation history: Modeling of visual binaries}",
	journal = {\na},
	year = "2018",
	month = "May",
	volume = {61},
	pages = {24-29},
	doi = {10.1016/j.newast.2017.11.005},
	adsurl = {https://ui.adsabs.harvard.edu/abs/2018NewA...61...24G}
}

@ARTICLE{2010MNRAS.404.1835K,
	author = {{Kouwenhoven}, M.~B.~N. and {Goodwin}, S.~P. and {Parker}, Richard J. and
	{Davies}, M.~B. and {Malmberg}, D. and {Kroupa}, P.},
	title = "{The formation of very wide binaries during the star cluster dissolution phase}",
	journal = {\mnras},
	year = "2010",
	month = "Jun",
	volume = {404},
	number = {4},
	pages = {1835-1848},
	doi = {10.1111/j.1365-2966.2010.16399.x},
	archivePrefix = {arXiv},
	eprint = {1001.3969},
	primaryClass = {astro-ph.GA},
	adsurl = {https://ui.adsabs.harvard.edu/abs/2010MNRAS.404.1835K}
}

@ARTICLE{2005A&A...439..565G,
	author = {{Goodwin}, S.~P. and {Kroupa}, P.},
	title = "{Limits on the primordial stellar multiplicity}",
	journal = {\aap},
	year = "2005",
	month = "Aug",
	volume = {439},
	number = {2},
	pages = {565-569},
	doi = {10.1051/0004-6361:20052654},
	archivePrefix = {arXiv},
	eprint = {astro-ph/0505470},
	primaryClass = {astro-ph},
	adsurl = {https://ui.adsabs.harvard.edu/abs/2005A&A...439..565G}
}

@ARTICLE{2013ARA&A..51..269D,
	author = {{Duch{\^e}ne}, Gaspard and {Kraus}, Adam},
	title = "{Stellar Multiplicity}",
	journal = {\araa},
	year = "2013",
	month = "Aug",
	volume = {51},
	number = {1},
	pages = {269-310},
	doi = {10.1146/annurev-astro-081710-102602},
	archivePrefix = {arXiv},
	eprint = {1303.3028},
	primaryClass = {astro-ph.SR},
	adsurl = {https://ui.adsabs.harvard.edu/abs/2013ARA&A..51..269D}
}

@ARTICLE{2014AJ....147...87T,
	author = {{Tokovinin}, Andrei},
	title = "{From Binaries to Multiples. II. Hierarchical Multiplicity of F and G Dwarfs}",
	journal = {\aj},
	year = "2014",
	month = "Apr",
	volume = {147},
	number = {4},
	eid = {87},
	pages = {87},
	doi = {10.1088/0004-6256/147/4/87},
	archivePrefix = {arXiv},
	eprint = {1401.6827},
	primaryClass = {astro-ph.SR},
	adsurl = {https://ui.adsabs.harvard.edu/abs/2014AJ....147...87T}
}

@ARTICLE{2016BaltA..25...49M,
	author = {{Malkov}, O. and {Karchevsky}, A. and {Kaygorodov}, P. and
	{Kovaleva}, D.},
	title = "{Identification list of binaries}",
	journal = {Baltic Astronomy},
	year = "2016",
	month = "Jan",
	volume = {25},
	pages = {49-52},
	doi = {10.1515/astro-2017-0109},
	adsurl = {https://ui.adsabs.harvard.edu/abs/2016BaltA..25...49M}
}

@ARTICLE{2001ApJ...555..945K,
	author = {{Kroupa}, Pavel and {Burkert}, Andreas},
	title = "{On the Origin of the Distribution of Binary Star Periods}",
	journal = {\apj},
	year = "2001",
	month = "Jul",
	volume = {555},
	number = {2},
	pages = {945-949},
	doi = {10.1086/321515},
	archivePrefix = {arXiv},
	eprint = {astro-ph/0103429},
	primaryClass = {astro-ph},
	adsurl = {https://ui.adsabs.harvard.edu/abs/2001ApJ...555..945K}
}

@ARTICLE{2009A&A...493..979K,
	author = {{Kouwenhoven}, M.~B.~N. and {Brown}, A.~G.~A. and {Goodwin}, S.~P. and
	{Portegies Zwart}, S.~F. and {Kaper}, L.},
	title = "{Exploring the consequences of pairing algorithms for binary stars}",
	journal = {\aap},
	year = "2009",
	month = "Jan",
	volume = {493},
	number = {3},
	pages = {979-1016},
	doi = {10.1051/0004-6361:200810234},
	archivePrefix = {arXiv},
	eprint = {0811.2859},
	primaryClass = {astro-ph},
	adsurl = {https://ui.adsabs.harvard.edu/abs/2009A&A...493..979K}
}

@ARTICLE{2001A&A...365..491N,
	author = {{Nelemans}, G. and {Yungelson}, L.~R. and {Portegies Zwart}, S.~F. and
	{Verbunt}, F.},
	title = "{Population synthesis for double white dwarfs . I. Close detached systems}",
	journal = {\aap},
	year = "2001",
	month = "Jan",
	volume = {365},
	pages = {491-507},
	doi = {10.1051/0004-6361:20000147},
	archivePrefix = {arXiv},
	eprint = {astro-ph/0010457},
	primaryClass = {astro-ph},
	adsurl = {https://ui.adsabs.harvard.edu/abs/2001A&A...365..491N}
}

@ARTICLE{2001MNRAS.322..231K,
	author = {{Kroupa}, Pavel},
	title = "{On the variation of the initial mass function}",
	journal = {\mnras},
	year = "2001",
	month = "Apr",
	volume = {322},
	number = {2},
	pages = {231-246},
	doi = {10.1046/j.1365-8711.2001.04022.x},
	archivePrefix = {arXiv},
	eprint = {astro-ph/0009005},
	primaryClass = {astro-ph},
	adsurl = {https://ui.adsabs.harvard.edu/abs/2001MNRAS.322..231K}
}

@ARTICLE{2016Natur.538..483T,
	author = {{Tobin}, John J. and {Kratter}, Kaitlin M. and {Persson}, Magnus V. and
	{Looney}, Leslie W. and {Dunham}, Michael M. and
	{Segura-Cox}, Dominique and {Li}, Zhi-Yun and {Chandler}, Claire J. and
	{Sadavoy}, Sarah I. and {Harris}, Robert J. and {Melis}, Carl and
	{P{\'e}rez}, Laura M.},
	title = "{A triple protostar system formed via fragmentation of a gravitationally unstable disk}",
	journal = {\nat},
	year = "2016",
	month = "Oct",
	volume = {538},
	number = {7626},
	pages = {483-486},
	doi = {10.1038/nature20094},
	archivePrefix = {arXiv},
	eprint = {1610.08524},
	primaryClass = {astro-ph.SR},
	adsurl = {https://ui.adsabs.harvard.edu/abs/2016Natur.538..483T}
}

@ARTICLE{2000MNRAS.315..543H,
	author = {{Hurley}, Jarrod R. and {Pols}, Onno R. and {Tout}, Christopher A.},
	title = "{Comprehensive analytic formulae for stellar evolution as a function of mass and metallicity}",
	journal = {\mnras},
	year = "2000",
	month = "Jul",
	volume = {315},
	number = {3},
	pages = {543-569},
	doi = {10.1046/j.1365-8711.2000.03426.x},
	archivePrefix = {arXiv},
	eprint = {astro-ph/0001295},
	primaryClass = {astro-ph},
	adsurl = {https://ui.adsabs.harvard.edu/abs/2000MNRAS.315..543H}
}

@ARTICLE{2015A&C....11..119K,
	author = {{Kovaleva}, D. and {Kaygorodov}, P. and {Malkov}, O.
	and {Debray}, B. and
	{Oblak}, E.},
	title = "{Binary star DataBase BDB development: Structure,
	algorithms, and VO standards implementation}",
	journal = {Astronomy and Computing},
	year = "2015",
	month = "Jun",
	volume = {11},
	pages = {119-125},
	doi = {10.1016/j.ascom.2015.02.007},
	adsurl =
	{https://ui.adsabs.harvard.edu/abs/2015A&C....11..119K}
}

@ARTICLE{2017ARep...61...80S,
	author = {{Samus'}, N.~N. and {Kazarovets}, E.~V. and {Durlevich},
	O.~V. and
	{Kireeva}, N.~N. and {Pastukhova}, E.~N.},
	title = "{General catalogue of variable stars: Version GCVS
	5.1}",
	journal = {Astronomy Reports},
	year = 2017,
	month = jan,
	volume = 61,
	pages = {80-88},
	doi = {10.1134/S1063772917010085},
	adsurl = {http://adsabs.harvard.edu/abs/2017ARep...61...80S}
}

@article{astropy:2013,
	Adsurl = {http://adsabs.harvard.edu/abs/2013A%26A...558A..33A},
	Archiveprefix = {arXiv},
	Author = {{Astropy Collaboration} and {Robitaille}, T.~P. and {Tollerud}, E.~J. and {Greenfield}, P. and {Droettboom}, M. and {Bray}, E. and {Aldcroft}, T. and {Davis}, M. and {Ginsburg}, A. and {Price-Whelan}, A.~M. and {Kerzendorf}, W.~E. and {Conley}, A. and {Crighton}, N. and {Barbary}, K. and {Muna}, D. and {Ferguson}, H. and {Grollier}, F. and {Parikh}, M.~M. and {Nair}, P.~H. and {Unther}, H.~M. and {Deil}, C. and {Woillez}, J. and {Conseil}, S. and {Kramer}, R. and {Turner}, J.~E.~H. and {Singer}, L. and {Fox}, R. and {Weaver}, B.~A. and {Zabalza}, V. and {Edwards}, Z.~I. and {Azalee Bostroem}, K. and {Burke}, D.~J. and {Casey}, A.~R. and {Crawford}, S.~M. and {Dencheva}, N. and {Ely}, J. and {Jenness}, T. and {Labrie}, K. and {Lim}, P.~L. and {Pierfederici}, F. and {Pontzen}, A. and {Ptak}, A. and {Refsdal}, B. and {Servillat}, M. and {Streicher}, O.},
	Doi = {10.1051/0004-6361/201322068},
	Eid = {A33},
	Eprint = {1307.6212},
	Journal = {\aap},
	Month = oct,
	Pages = {A33},
	Primaryclass = {astro-ph.IM},
	Title = {{Astropy: A community Python package for astronomy}},
	Volume = 558,
	Year = 2013}

@article{astropy:2018,
	Adsurl = {https://ui.adsabs.harvard.edu/#abs/2018AJ....156..123T},
	Author = {{Price-Whelan}, A.~M. and {Sip{\H{o}}cz}, B.~M. and {G{\"u}nther}, H.~M. and {Lim}, P.~L. and {Crawford}, S.~M. and {Conseil}, S. and {Shupe}, D.~L. and {Craig}, M.~W. and {Dencheva}, N. and {Ginsburg}, A. and {VanderPlas}, J.~T. and {Bradley}, L.~D. and {P{\'e}rez-Su{\'a}rez}, D. and {de Val-Borro}, M. and {Paper Contributors}, (Primary and {Aldcroft}, T.~L. and {Cruz}, K.~L. and {Robitaille}, T.~P. and {Tollerud}, E.~J. and {Coordination Committee}, (Astropy and {Ardelean}, C. and {Babej}, T. and {Bach}, Y.~P. and {Bachetti}, M. and {Bakanov}, A.~V. and {Bamford}, S.~P. and {Barentsen}, G. and {Barmby}, P. and {Baumbach}, A. and {Berry}, K.~L. and {Biscani}, F. and {Boquien}, M. and {Bostroem}, K.~A. and {Bouma}, L.~G. and {Brammer}, G.~B. and {Bray}, E.~M. and {Breytenbach}, H. and {Buddelmeijer}, H. and {Burke}, D.~J. and {Calderone}, G. and {Cano Rodr{\'\i}guez}, J.~L. and {Cara}, M. and {Cardoso}, J.~V.~M. and {Cheedella}, S. and {Copin}, Y. and {Corrales}, L. and {Crichton}, D. and {D{\textquoteright}Avella}, D. and {Deil}, C. and {Depagne}, {\'E}. and {Dietrich}, J.~P. and {Donath}, A. and {Droettboom}, M. and {Earl}, N. and {Erben}, T. and {Fabbro}, S. and {Ferreira}, L.~A. and {Finethy}, T. and {Fox}, R.~T. and {Garrison}, L.~H. and {Gibbons}, S.~L.~J. and {Goldstein}, D.~A. and {Gommers}, R. and {Greco}, J.~P. and {Greenfield}, P. and {Groener}, A.~M. and {Grollier}, F. and {Hagen}, A. and {Hirst}, P. and {Homeier}, D. and {Horton}, A.~J. and {Hosseinzadeh}, G. and {Hu}, L. and {Hunkeler}, J.~S. and {Ivezi{\'c}}, {\v{Z}}. and {Jain}, A. and {Jenness}, T. and {Kanarek}, G. and {Kendrew}, S. and {Kern}, N.~S. and {Kerzendorf}, W.~E. and {Khvalko}, A. and {King}, J. and {Kirkby}, D. and {Kulkarni}, A.~M. and {Kumar}, A. and {Lee}, A. and {Lenz}, D. and {Littlefair}, S.~P. and {Ma}, Z. and {Macleod}, D.~M. and {Mastropietro}, M. and {McCully}, C. and {Montagnac}, S. and {Morris}, B.~M. and {Mueller}, M. and {Mumford}, S.~J. and {Muna}, D. and {Murphy}, N.~A. and {Nelson}, S. and {Nguyen}, G.~H. and {Ninan}, J.~P. and {N{\"o}the}, M. and {Ogaz}, S. and {Oh}, S. and {Parejko}, J.~K. and {Parley}, N. and {Pascual}, S. and {Patil}, R. and {Patil}, A.~A. and {Plunkett}, A.~L. and {Prochaska}, J.~X. and {Rastogi}, T. and {Reddy Janga}, V. and {Sabater}, J. and {Sakurikar}, P. and {Seifert}, M. and {Sherbert}, L.~E. and {Sherwood-Taylor}, H. and {Shih}, A.~Y. and {Sick}, J. and {Silbiger}, M.~T. and {Singanamalla}, S. and {Singer}, L.~P. and {Sladen}, P.~H. and {Sooley}, K.~A. and {Sornarajah}, S. and {Streicher}, O. and {Teuben}, P. and {Thomas}, S.~W. and {Tremblay}, G.~R. and {Turner}, J.~E.~H. and {Terr{\'o}n}, V. and {van Kerkwijk}, M.~H. and {de la Vega}, A. and {Watkins}, L.~L. and {Weaver}, B.~A. and {Whitmore}, J.~B. and {Woillez}, J. and {Zabalza}, V. and {Contributors}, (Astropy},
	Doi = {10.3847/1538-3881/aabc4f},
	Eid = {123},
	Journal = {\aj},
	Month = Sep,
	Pages = {123},
	Primaryclass = {astro-ph.IM},
	Title = {{The Astropy Project: Building an Open-science Project and Status of the v2.0 Core Package}},
	Volume = {156},
	Year = 2018}
\end{document}